\documentclass[journal,11pt,draftclsnofoot,onecolumn]{IEEEtran}
\usepackage{geometry} 
\usepackage[parfill]{parskip}    
\usepackage{graphicx}
\usepackage{amssymb}
\usepackage{epstopdf}
\usepackage{subfigure}
      \makeatletter
      
      \@addtoreset{equation}{section}
      \makeatother

\newtheorem{teiri}{Theorem}[section]
\newtheorem{kei}{Corollary}[section]

\newtheorem{hodai}{Lemma}[section]
\newtheorem{teigi}{Definition}[section]
\newtheorem{rei}{Example}[section]
\newtheorem{chui}{Remark}[section]

\usepackage{type1cm}

\newcommand{\Exp}{{\rm E}}

\newcommand{\spr}[1]{{\bf #1}}

\newcommand{\vep}{\varepsilon}
\newcommand{\vph}{\varphi}

\newcommand{\cC}{{\cal C}}

\newcommand{\cM}{{\cal M}}

\newcommand{\cU}{{\cal U}} 
\newcommand{\cV}{{\cal V}}
\newcommand{\cX}{{\cal X}}
\newcommand{\cY}{{\cal Y}}
\newcommand{\cT}{{\cal T}}
\newcommand{\cZ}{{\cal Z}}

\newcommand{\sR}{\spr{R}}

\newcommand{\ssc}{\spr{c}}

\newcommand{\ssv}{\spr{v}} 
\newcommand{\ssx}{\spr{x}}
\newcommand{\ssz}{\spr{z}}
\newcommand{\ssy}{\spr{y}}

\newcommand{\nth}{\frac{1}{n}}

\newcommand{\lin}{\liminf_{n \to \infty}}

\newcommand{\bteiri}{\begin{teiri}}
\newcommand{\eteiri}{\end{teiri}}
\newcommand{\bkei}{\begin{kei}}
\newcommand{\ekei}{\end{kei}}
\newcommand{\brei}{\begin{rei}}
\newcommand{\erei}{\end{rei}}
\newcommand{\bhodai}{\begin{hodai}}
\newcommand{\ehodai}{\end{hodai}}
\newcommand{\bteigi}{\begin{teigi}}
\newcommand{\eteigi}{\end{teigi}}
\newcommand{\bchui}{\begin{chui}}
\newcommand{\echui}{\end{chui}}
\newcommand{\beq}{\begin{equation}}
\newcommand{\eeq}{\end{equation}}
\newcommand{\beqn}{\begin{eqnarray}}
\newcommand{\eeqn}{\end{eqnarray}}
\newcommand{\beqns}{\begin{eqnarray*}}
\newcommand{\eeqns}{\end{eqnarray*}}

\newcommand{\map}{\vph_n: \cX^n \to \cY^n}
\newcommand{\mapMtoY}{\vph_n: \cM_{M_n} \to \cY^n}
\newcommand{\mapXtoM}{\vph_n: \cX^n \to \cM_{M_n}}

\title{~\\~\\~\\~\\  Reliability and Secrecy Functions
of the Wiretap Channel under Cost Constraint}
\author{~\\~\\~\\~\\~Te~Sun~Han,~\IEEEmembership{Life Fellow,~IEEE},
\thanks{T. S. Han is  with the
Quantum ICT Laboratory, National Institute of Information and
Communications Technology (NICT), Nukui-kitamachi 4-2-1, Koganei,
Tokyo,184-8795, Japan (email: han@is.uec.ac.jp, han@nict.go.jp)} 
\  \ Hiroyuki Endo,\thanks{H. Endo is with the Department of Applied Physics, Waseda University,
Okubo 3-4-1, Shinjuku, Tokyo, Japan, and is also a
collaborating research fellow of the Quantum ICT Laboratory, NICT
 (email: h-endo-1212@ruri.waseda.jp, h-endo@nict.go.jp)}
\ \ Masahide Sasaki\thanks{M. Sasaki is  with the
Quantum ICT Laboratory, NICT, Nukui-kitamachi 4-2-1, Koganei,
Tokyo,184-8795, Japan (email: psasaki@nict.go.jp)}
}

\date{\today}

\begin{document}
\setcounter{page}{0}
\maketitle
\thispagestyle{empty}
\newpage

\pagenumbering{roman}

%
%

\pagenumbering{arabic}
\setcounter{page}{1}
\setcounter{equation}{0}

%


\begin{abstract} The wiretap channel has been devised and studied first by Wyner, and subsequently 
extended to the case with non-degraded general wiretap channels by Csisz\'ar and K\"orner.
Focusing mainly on  the stationary memoryless channel with cost constraint, we  newly introduce the notion 
of reliability and secrecy functions as a fundamental tool 
to analyze and/or design the performance of an efficient wiretap channel system,
including binary symmetric  wiretap channels, Poisson wiretap channels and Gaussian wiretap channels.
Compact formulae for those functions are explicitly given for stationary memoryless wiretap channels.
It is also demonstrated that,
based on such a {\em pair} of reliability and secrecy functions, we can control the tradeoff between 
reliability and secrecy (usually conflicting), both with exponentially decreasing rates as block length $n$
 becomes large.
Four ways to  do so are given on the basis of  rate shifting,  rate exchange, concatenation and change of cost constraint. 
Also,
the notion of the $\delta$ secrecy capacity is defined and shown to attain the strongest secrecy standard
among others. The maximized vs. averaged secrecy measures is also discussed.  
\end{abstract}



%

\section{Introduction}\label{introduction1}

The pioneering work by Wyner 
\cite{wyner-wire} as well as by
 Csisz\'ar and K\"orner 
\cite{csis-kor-3rd}, 
based on the wiretap channel model, 
has provided a strong impetus to find a new scheme 
of the physical layer cryptography 
in a good balance of usability and secrecy. 
In particular, they have  first formulated the tradeoff between the transmission rate for Bob 
and the 
equivocation rate against Eve.
Since then, 
``information theoretic security attracts much attention, because it offers security that does not depend 
on conjectured difficulties of some computational problem, "
\footnote{suggested by Associate Editor}
 and
there have been extensive studies 
on various kinds of wiretap channels, 
which are nicely summarized, e.g.,  in Laourine and  Wagner \cite{wagner} along with
 the secrecy capacity formula for the Poisson wiretap channel  
without cost constraint.
Among others, Hayashi \cite{hayashi-exp} is the first who has derived 
the relevant secrecy exponent function to specify the exponentially decreasing speed
 (i.e., exponent) of the leaked information 
under the {\em average  secrecy criterion}
when
{\em no cost constraint} is considered.
\\
\ \
Throughout in this paper, we are imposed   {\em cost constraints}
(limit on available transmission energy, bandwidth, and so on).
We first address, given a general wiretap channel,  the primal problem to establish a general formula to simultaneously 
summarize the reliability performance for Bob 
and the secrecy performance against Eve under the {\em maximum  secrecy criterion}.
Next, it is shown that  both of them are described by using exponentially decaying 
 functions of the code length when a stationary memoryless wiretap channel is considered.
 This provides the theoretical basis for investigating the asymptotic behavior of reliability and secrecy.
We can then specifically quantify achievable reliability  exponents and achievable  secrecy exponents
 as well as the tradeoff between them
 for several important wiretap channel models
such as binary symmetric wiretap channels, Poisson wiretap channels, Gaussian wiretap channels.  
In particular, four  ways of the tradeoff to control reliability and secrecy are given and discussed with their novel significance.
Also, on the basis of the analysis of these exponents under cost constraint, 
the new  formula for the $\delta$-secrecy capacity (with the strongest secrecy among others) is established
to apply to several typical wiretap channel models. 
A  remarkable feature of this paper is that we first derive the key formulas  not depending on respective specific 
channel models and then apply them 
 to those respective cases
 to get new insights into each case
 as well.
\\
\ \ 
The paper is organized as follows.  
In Section \ref{intro-geri1}, the definitions of  {\em wiretap channel} and related notions
such as {\em error probability},  {\em cost constraint}, {\em secrecy capacity} and {\em concatenation} 
are introduced along with various kinds of 
{\em secrecy measures}. 
\\
\ \ 
In Section \ref{intro-geri2}.A, we give 
a fundamental formula to simultaneously evaluate a {\em pair} of  
 reliability behavior and secrecy behavior  under  cost constraint for a {\em general} wiretap channel, which is then
 in Section \ref{intro-geri2}.B,
 particularized to establish the specific formulas for 
 stationary and memoryless  wiretap channels. 
 Here, the notions of  {\em reliability function} and {\em secrecy function} are introduced  to evaluate
 the exponent of the exponentially decreasing  decoding error for Bob and that of the exponentially 
 decreasing divergence distance against Eve
 for the {\em stationary memoryless} wiretap channel under cost constraint.
 This is one of the key results in this paper. 
We also present their numerical examples to see how the reliability and secrecy exponents 
vary depending on the channel and cost parameters. 
Also,  superiority of the maximum secrecy criterion to the average secrecy criterion is discussed.
In Section \ref{intro-geri2}.C, a strengthening  of Theorem \ref{teiri:hanw2} in Section \ref{intro-geri2}.B is provided.
In Section \ref{intro-geri2}.D, the $\delta$-secrecy capacity formula 
(with the strongest secrecy) is given under cost constraint, including the formula for a special but important case
 with {\em more capable} wiretap channels. 
\\
\ \ 
In Section \ref{sec:tradeoff},
four ways for the tradeoff are demonstrated: one is 
by rate shifting, another one  by rate exchange,
one more by concatenation,  and the other by
change of cost constraint, which are discussed in terms of the reliability and secrecy exponents.
This section is  thus prepared for more quantitative analysis/design of the reliability-secrecy tradeoff. 
\\
\ \ 
In Section \ref{intro-poisson}, the formula for the $\delta$-secrecy capacity is applied to the Poisson wiretap channel with cost constraint, which is a practical model for free-space Laser communication with a photon counter. 
\\
\ \ 
In Section \ref{intro-poisson-re-se2}, for Poisson wiretap channels with cost constraint we demonstrate 
the reliability and secrecy functions as an application of the key theorem established in Section \ref{intro-geri2}.B.
\\
\ \ 
In Section 
\ref{intro-poisson-re-conc1}, we investigate the effects of channel concatenation 
with an auxiliary channel for the Poisson wiretap channel. 
\\
\ \ 
In Section \ref{intro-gaussian-re-se2}, the $\delta$-secrecy capacity formula for the Gaussian wiretap channel
is given as an application of the key theorem established in Section \ref{intro-geri2}.D.
\\
\ \ 
In Section \ref{intro-gaussian-re-se4}, for the Gaussian wiretap channels with cost constraint we demonstrate 
the reliability and secrecy functions as an application of the key theorem established in Section \ref{intro-geri2}.B.
In particular, these functions are numerically compared with those of Gallager-type, which reveals that
a kind of duality exists among them.
In Section \ref{conc-remark}, we conclude the paper.

\section{Preliminaries and basic concepts}\label{intro-geri1}

%
%
In this section we give the definition of the wiretap channel. There are
several levels and ways to specify the superiority of the legitimate users,
Alice and Bob, to the eavesdropper, Eve, such as physically degraded 
Eve, (statistically) degraded Eve, less noisy  Bob,  and more capable
Bob. In this paper, we are interested mainly in  the last class of channels because 
the other ones
 imply the last one (cf. Csisz\'ar and K\"orner \cite{csis-kor-2nd}).
\\
\ \
We  introduce here the necessary notions and  notations to quantify the reliability and the
secrecy of this kind of wiretap channel model. In particular, we define
several kinds of secrecy metrics, including the strongest criterion based
on the divergence distance with reference to a target output distribution, while
the notion of concatenation of channels is also introduced to construct a possible way to
control tradeoff 
between reliability and  secrecy.
%
%
%





{\em A. Wiretap channel}

Let $\cX, \cY, \cZ$ be arbitrary alphabets (not necessarily finite),
where $\cX$ is called an {\em input alphabet}, and $\cY, \cZ$ are called
{\em output alphabets.}
A general {\em wiretap channel} consists of two  general channels, i.e.,  
$W_B^{n}: \cX^n\to\cY^n$
 (from Alice for Bob) and  
$W_E^{n}: \cX^n\to\cZ^n$
 (from Alice against Eve), where 
$W_B^{n}(\ssy|\ssx)$, $W_E^{n}(\ssz|\ssx)$ are the conditional probabilities of
$\ssy \in \cY^n, \ssz \in \cZ^n$ given $\ssx \in \cX^n$ (of block length $n$), respectively.
Alice wants to communicate with Bob as reliably as possible but  
as secretly as possible against Eve.
We let  ($W_B^{n}, W_E^{n})$ indicate such a wiretap channel.
\\
\ \
Given  a message set $\cM_n \equiv \{1,2,\cdots, M_n\}$, we consider 
a {\em  stochastic} encoder  for Alice
             $\varphi_n: \cM_n\to \cX^n$ and a decoder for Bob $\psi_n^{B}:\cY^n\to \cM_n$,
             and  for $i\in \cM_n$  let
             $\varphi_n^B(i)$ denote the output due to $\varphi_n(i)$ via channel $W_B^n$.
             
 {\em B. Cost constraint}     
 
 From the viewpoint of communication technologies, 
it is sometimes needed to impose {\em  cost constraint} on channel inputs. 
Here we give its formal definition.
\\
\ \
For $n=1, 2,\cdots$ fix a mapping $c_n: \cX^n\to {\bf R}^+$ (the set of nonnegative real numbers) arbitrarily. 
For $\ssx\in \cX^n$ we call $c_n(\ssx)$ the cost of $\ssx$ and 
$\frac{1}{n}c_n(\ssx)$ the cost per letter. In the channel coding problem 
with cost constraint, we require  the 
encoder outputs 
$\varphi_n(i) \in \cX^n$
satisfy 
\beq\label{eq:w-cost-per}
\Pr\left\{\frac{1}{n}c_n(\varphi_n(i)) \le \Gamma\right\}=1\quad (\mbox{for all\ }i=1,2,\cdots, M_n), 
\eeq
where $\Gamma$ is an arbitrarily nonnegative given constant, 
which we call {\em cost constraint} $\Gamma$. 
Notice here that the encoder $\varphi_n$ is {\em stochastic}.   
 When  
(\ref{eq:w-cost-per}) holds,
we say that the encoder $\varphi_n$ satisfies the cost constraint $\Gamma$ and call 
 ($W_B^{n}, W_E^{n})$ a wiretap channel with {\em cost constraint} $\Gamma$.
Incidentally, define 
   \beq\label{eq:wag-c1}
\cX^n(\Gamma) = \left\{ \ssx \in \cX^n\left|\frac{1}{n}c_n(\ssx) \le \Gamma\right.\right\},
\eeq
then (\ref{eq:w-cost-per}) is rewritten also as 
\beq\label{eq:wag-we1}
\Pr\left\{\varphi_n(i) \in \cX^n(\Gamma)\right\}  = 1\quad (\mbox{for all\ }i=1,2,\cdots, M_n).
\eeq                   
\bchui\label{chui:wagc1}
{\rm
Consider the case with $c_n(\ssx) =n\ (\forall \ssx \in \cX^n)$ and $\Gamma=1$, 
then in this case it is easy to check that
$\cX^n(\Gamma) =\cX^n$, which means that the wiretap channel is actually imposed no cost constraint.
} \QED
\echui

{\em C.  Error probability, secrecy measures and secrecy capacities}

            Given a wiretap channel ($W_B^{n}, W_E^{n}$) with cost constraint $\Gamma$, 
            the error probability $\epsilon_n^{B}$ ({\em measure of reliability}) 
            via channel $W_B^n$ 
            for Bob is defined to be 
         \beq\label{eq:memo1}
          \epsilon_n^{B}\equiv \frac{1}{M_n}\sum_{i \in \cM_n}
          \Pr\left\{\psi_n^{B}(\varphi_n^B(i))\neq i\right\},
           \eeq
             whereas  the {\em divergence distance} ({\em measure 1 of secrecy}) $ \delta_n^{E}$ and the
             {\em variational distance}  ({\em measure 2 of secrecy}) $\partial_n^{E}$
              via channel $W^n_E$ against Eve are defined to be 
            %
            \beq\label{eq:memo2}
             \delta_n^{E}\equiv \frac{1}{M_n}
            \sum_{i\in \cM_n}D(P^{(i)}_n||\pi_n),
           \eeq
           \beq\label{eq:memo2comp}
             \partial_n^{E}\equiv \frac{1}{M_n}
            \sum_{i\in \cM_n}d(P^{(i)}_n, \pi_n)
           \eeq
          where                          \[
             D(P_1||P_2) =\sum_{u\in \cU}P_1(u)\log\frac{P_1(u)}{P_2(u)},
             \]
              \[
             d(P_1, P_2) =\sum_{u\in \cU}|P_1(u)-P_2(u)|;
               \]
            where             %
            $P^{(i)}_n$ denotes the output probability distribution on $\cZ^n$ 
            via  channel $W_E^{n}$ due to the input 
            $\varphi_n(i)$, and 
            $\pi_n$ is called the {\em target} output  probability distribution on $\cZ^n$,
            which is generated   
            via channel $W_E^{n}$ due to an arbitrarily  prescribed input distribution on $\cX^n$.   
            Specifically, $\pi_n$ is given by $\pi_n(\ssz) = \sum_{\ssx\in \cX^n}W^n(\ssz|\ssx)P_{X^n}(\ssx)$.
            In this paper the logarithm is taken to the natural base $e$.         %
               \\
            \ \ 
             With these two typical measures of secrecy, 
             we can define two kinds of criteria for {\em achievability}:
                 \beq\label{eq:memo3conq}
             \epsilon_n^{B}\to 0,\  \delta_n^{E}\to 0\quad  \mbox{as $n\to\infty$},
             \eeq
                 \beq\label{eq:memo3}
             \epsilon_n^{B}\to 0,\  \partial_n^{E}\to 0\quad  \mbox{as $n\to\infty$}.
             \eeq
              We say that a rate $R$ is {\em $(\delta, \Gamma)$-achievable}  
               if there exists a pair $(\varphi_n, \psi_n^B)$
             of encoder and decoder satisfying criterion (\ref{eq:memo3conq}) and 
           \beq\label{eq:cost-nasi}
             \liminf_{n\to\infty}\nth\log M_n \ge R.
          \eeq
          When there is no fear of confusion, we say simply that a rate $R$ is
          {\em $\delta$-achievable} by dropping cost constraint $\Gamma$, and so on also in the sequel.
          Similarly, we say that a rate $R$ is {\em $(\partial, \Gamma)$-achievable}
               if there exists a pair $(\varphi_n, \psi_n^B)$
             of encoder and decoder satisfying criterion (\ref{eq:memo3}) and (\ref{eq:cost-nasi}).
             It should be noted here that criterion (\ref{eq:memo3conq}) implies criterion (\ref{eq:memo3}),
             owing to Pinsker inequality \cite{pinsker}:
             \[
             \left(\partial_n^{E}\right)^2 \le 2\delta_n^{E},
             \]
             which means that criterion (\ref{eq:memo3conq}) is stronger than criterion (\ref{eq:memo3}).
             %
             %
            %
            
\begin{figure}[htbp]
\begin{center}
\includegraphics[width=50mm]{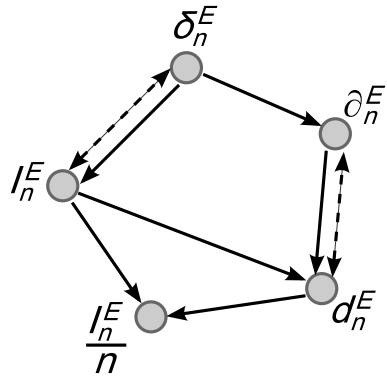}
\end{center}
\caption{The implication scheme: The arrow $\alpha \longrightarrow \beta$
 means that $\alpha$ is stronger than $\beta$;
         $\alpha \dashleftarrow\dashrightarrow \beta$ means that $\alpha$ coincides with $\beta$
         when $\pi_n=P_n$, where $I^E_n \rightarrow d^E_n$ is due to \cite{pinsker} and 
         $d^E_n \rightarrow \frac{1}{n}I^E_n$
          is due to \cite{bloch}.  In the finite alphabet case,
          exponential decay of $d_n^E$ (with increasing $n$) implies that of  $I_n^E$
          (cf. \cite{csis-all}).
          }
\label{fig0}
\end{figure}

                         On the other hand,  many people (e.g., Csisz\'ar \cite{csis-all},  Hayashi \cite{hayashi-exp})
             have used, instead of
              measure (\ref{eq:memo2}), 
               the {\em mutual information}:
               \beq\label{eq:criter-inform}
               I_n^E \equiv \frac{1}{M_n}\sum_{i\in \cM_n}D(P^{(i)}_n||P_n),
               \quad P_n = \frac{1}{M_n}\sum_{i\in \cM_n}P^{(i)}_n.
               \eeq  
               With this measure ({\em measure 3 of secrecy}), 
               we may consider one more criterion for achievability (called  the i-achievability):
                   \beq\label{eq:memo3conq-1}
             \epsilon_n^{B}\to 0,\  I_n^{E}\to 0\quad  \mbox{as $n\to\infty$}.
             \eeq
                 On the other hand, since the identity (Pythagorean theorem):
                \beq\label{eq:wagc2}
         \delta_n^{E} =I_n^E + D(P_n ||\pi_n) 
        \eeq
         holds, 
              $\delta_n^{E}$ is a stronger measure  than $I_n^E$.
               Moreover, since
               \[
              d_n^E \equiv \frac{1}{M_n}\sum_{i\in \cM_n}
              d(P_n^{(i)}, P_n) \le \frac{2}{M_n}\sum_{i\in \cM_n}d(P_n^{(i)}, \pi_n) = 2\partial_n^E
               \]
               always holds by virtue of the triangle axiom of the variational distance,
               $\partial_n^E$ is stronger than $d_n^E$  ({\em measure 4 of secrecy}: 
               cf. \cite{csis-all}), so that criterion
                (\ref{eq:memo3}) is stronger than the d-achievability:
               \beq\label{eq:memo3conq-21}
             \epsilon_n^{B}\to 0,\  d_n^{E}\to 0\quad  \mbox{as $n\to\infty$}.
             \eeq
               Furthermore, one may sometimes prefer to consider 
               the following achievability (called the w-achievability):
               \beq\label{eq:memo3conq-2}
             \epsilon_n^{B}\to 0,\  \nth I_n^{E}\to 0\quad  \mbox{as $n\to\infty$},
             \eeq
             which is nothing but  the so-called  weak secrecy ({\em measure 5 of secrecy}).
             Indeed, this is the {\em weakest} criterion among others; its illustrating
              example will appear 
             in Examples \ref{rei:wag5} and \ref{rei:gaussq1}, while criterion (\ref{eq:memo3conq})
              is the {\em strongest} one and introduced for the first time in this paper.
             Fig.\ref{fig0} shows the implication scheme among these five measures of secrecy.
             \\
             \ \ 
             The secrecy capacities $\delta$-$C_s(\Gamma) $ and $\partial\mbox{-}C_s(\Gamma)$ 
             between Alice and Bob are defined
              to be the supremum of  all  $(\delta, \Gamma)$-achievable rates and that of  
               all $(\partial, \Gamma)$-achievable rates,
              respectively. Similarly, the secrecy capacity d-$C_s(\Gamma)$ with d-achievability,
              the secrecy capacity i-$C_s(\Gamma)$ with i-achievability as well as
               the secrecy capacity
              w-$C_s(\Gamma)$ with w-achievability can also be defined.
             \bchui\label{chui:wagc2A}
             {\rm  One may wonder if the ``strongest" measure $\delta^E_n$ of secrecy can be given an operational meaning.
             In this connection, we would like to cite the paper by Hou and Kramer \cite{hou-kra} in which
             $I_n^E$ is interpreted as a measure of ``non-confusion" and $D(P_n||\pi_n)$ 
             as a measure  of ``non-stealth," and $\pi_n$ is interpreted as the background noise distribution on $\cZ^n$ that 
             Eve detects in advance to the communication between Alice and Bob; thus, 
              in view of (\ref{eq:wagc2}), by making $\delta_n^E \to 0$ 
             we can not only keep the message secret from Eve but also hide the presence of meaningful communication.
             Alice can control $\pi_n$ so as to be most perplexng to Eve.
            A connection to some hypothesis testing problem is also pointed out. 
            A similar interpretation is given also for $\partial_n^E$  with 
            $d_n^E$ as a measure of  ``non-confusion" and $d(P_n, \pi_n)$ as a measure of  ``non-stealth,"
           because   the following inequality holds:
     \beq\label{eq:varwagh1}
           d_n^E + d(P_n, \pi_n) \le 3\partial_n^E.
         \eeq
            }
            \echui
            \bchui\label{chui:wagc2}
             {\rm  We notice  that all of  $\epsilon_n^{B}$, $\delta_n^{E}$,  
                 $\partial_n^E$, $d_n^E$  and $I_n^E$, $\nth I_n^E$
                 defined here are  the measures {\em averaged} over the message set $\cM_n$
                with  the uniform distribution.
            On the other hand, we can consider also the criteria {\em maximized} 
                                 over the message set $\cM_n,$
            which will be discussed later in Remark \ref{chui:max1}.
             \QED
            }
            \echui
            


{\em D. Concatenation}

In wiretap channel coding it is one of the important problems how to control  the tradeoff between
the reliability for Bob and the secrecy against Eve. There are several ways to control it.
One of these is to make use of  the {\em concatenation} of the main wiretap channel with an auxiliary (virtual)
channel. So, it is convenient to state here  its formal definition for later use.

Let $\cV$ be an arbitrary alphabet (not necessarily finite) and let $V^n$ be an arbitrary auxiliary 
 random variable with values in $\cV^n$ such that 
$V^n \to X^n\to Y^nZ^n$ forms a Markov chain in this order, 
where $X^n$ is  an input variable for the wiretap channel $(W_B^n, W_E^n)$; and 
$Y^n, Z^n$  are the output variables of channels $W^n_B, W^n_E$ due to the input $X^n$, respectively.
%
%
\bteigi\label{teigi:conc}
{\rm 
Given a general  channel $W^n: \cX^n \to \cY^n$, we define its {\em concatenated} channel
$W^{n+}: \cV^n \to \cY^n$ so that
\beq\label{eq:conc1}
W^{n+}(\ssy|\ssv)=\sum_{\ssx \in \cX^n}W^n(\ssy|\ssx)P_{X^n|V^n}(\ssx|\ssv),
\eeq
where
\footnote{We use the convention that, given random variables $S$ and $T$, $P_S(\cdot)$ and
$P_{S|T}(\cdot |\cdot)$ denote the probability distribution of $S$, and the conditional 
probability distribution of $S$ given $T$, respectively}
 $P_{X^n|V^n}: \cV^n\to \cX^n$ is an arbitrary auxiliary channel.
In particular, we say that  a pair $(W_B^{n+}, W_E^{n+})$ is a {\em concatenation} of 
the wiretap channel $(W_B^{n}, W_E^{n})$, if 
\beqn
W_B^{n+}(\ssy|\ssv)&=&\sum_{\ssx \in \cX^n}W_B^n(\ssy|\ssx)P_{X^n|V^n}(\ssx|\ssv)\label{eq:conc-cost1}
\\
W_E^{n+}(\ssz|\ssv)&=&\sum_{\ssx \in \cX^n}W_E^n(\ssz|\ssx)P_{X^n|V^n}(\ssx|\ssv). \label{eq:conc-cost2}
\eeqn
with the auxiliary channel $P_{X^n|V^n}$. Notice that if $V^n \equiv X^n$ as random variables then these  reduce to
the non-concenated wiretap channel.  
} \QED
\eteigi

{\em E. Stationary memoryless wiretap channel}

In this paper the substantial attention is payed to the  special class of wiretap channels
called the stationary memoryless wiretap channel, the definition of which is given by

\bteigi\label{teigi:wagc4}
{\rm A wiretap  channel $(W_B^{n}, W_E^{n})$  is said to be stationary and memoryless if,
with some channels $W_B:\cX\to \cY, W_E: \cX\to \cZ$, it holds that 
\beq\label{eq:wire1}
W_B^{n}(\ssy|\ssx) = \prod_{k=1}^n W_B(y_k|x_k), \quad W_E^n(\ssz|\ssx) = \prod_{k=1}^n W_E(z_k|x_k),
\eeq
where $\ssx =(x_1, x_2, \cdots, x_n),$ $\ssy =(y_1, y_2, \cdots, y_n),$ $\ssz =(z_1, z_2, \cdots,z_n).$
This wiretap channel may be denoted simply by $(W_B, W_E)$.
%
%
}
\QED
\eteigi
When we are dealing with a stationary memoryless wiretap channel $(W_B, W_E)$ it is usual to assume an
 {\em additive cost} $c: $ $\cX\to \sR^+$
in the sense that $c_n(\ssx) = \sum_{i=1}^{n}c(x_i)$ where $\ssx=(x_1, \cdots, x_n)$.
This enables us to analyze the detailed  performances of the wiretap channel, to be shown in the following sections.
%



\section{Evaluation of reliability  and secrecy  }\label{intro-geri2}

In this section, the problem of a general wiretap channel with general cost constraint $\Gamma$ is first studied, and 
next the problem of a stationary memoryless wiretap channel with additive cost constraint $\Gamma$ is
investigated in details. In particular, 
with  criterion (\ref{eq:memo3conq}) we are  interested in 
exponentially decreasing rates of $ \epsilon_n^{B}, \delta_n^{E}$ as $n$ tends to $\infty$.
Finally, its applicantion to establish a general formula for the $\delta$-secrecy capacity $\delta$-$C_s(\Gamma)$
with cost constraint is provided.

{\em A. General wiretap channel with cost constraint}

Let
$W^n(\ssy|\ssv): \cV^n\to \cY^n$, $W^n(\ssz|\ssv): \cV^n\to \cZ^n$
 be  arbitrary  general channels and $Q(\ssv)$ be an arbitrary  
auxiliary input distribution on $\cV^n$, and set
\beqn
\phi(\rho|W^{n}, Q) &\equiv & -\log\sum_{\ssy}\left(\sum_{\ssv}Q(\ssv)
W^{n}(\ssy|\ssv)^{\frac{1}{1+\rho}}\right)^{1+\rho},\label{eq:han1}\\
\psi(\rho|W^{n}, Q) &\equiv &
 -\log\sum_{\ssz}\left(\sum_{\ssv}Q(\ssv)W^{n}(\ssz|\ssv)^{1+\rho}\right)W^{n}_Q(\ssz)^{-\rho},\label{eeqhan2}
\eeqn
{\rm
where  $W_Q^{n}(\ssz) 
= \sum_{\ssv}Q(\ssv)W^{n}(\ssz|\ssv)$.  Then, we have
}
\bteiri \label{teiri:han2}
{\rm  Let $(W^n_B, W^n_E)$ be a general wiretap channel with general cost constraint $\Gamma$,  and 
$M_n$, $L_n$ be  arbitrary positive integers,
then there exists a pair 
$(\varphi_n, \psi_n^B$) of encoder (satisfying cost constraint $\Gamma$) and decoder such that 
\beqn
\epsilon_n^B & \le & 2\inf_{0\le \rho \le 1}(M_nL_n)^{\rho}e^{-\phi (\rho |W_B^{n+}, Q)},\label{eq:hana1}\\
\delta_n^E & \le &2\inf_{0<\rho \le1}\frac{e^{-\psi (\rho |W_E^{n+}, Q)}}{\rho L_n^{\rho}}\label{eq:hana2c}\\
& \le& 2\inf_{0<\rho <1}\frac{e^{-\phi (-\rho |W_E^{n+}, Q)}}{\rho L_n^{\rho}},\label{eq:hana2c2}
\eeqn
where 
$(W_B^{n+}, W_E^{n+})$ is a  concatenation of $(W_B^{n}, W_E^{n})$ 
 (cf. Definition \ref{teigi:conc}),
 and we assume that the condition 
 \beq\label{eq:wag-re1}
 \Pr\{X^n \in \cX^n(\Gamma)\}=1
 \eeq
 holds for
  the random variable $X^n$ over $\cX^n$ induced via the auxiliary channel $P_{X^n|V^n}$ 
   by the input variable $V^n$ subject to $Q(\ssv)$ on $\cV^n$.
 \QED
%
}
\eteiri
{\em Proof}: See Appendix \ref{appA}.
\bchui\label{chui:mutual-div1}
{\rm
Formula (\ref{eq:hana1}) {\em without} concatenation is due to Gallager \cite{gall}, while 
formulas (\ref{eq:hana2c}), (\ref{eq:hana2c2}) {\em without} concatenation and cost constraint 
 have first been shown  in a different context by Han and Verd\'u \cite[p.768]{ver-han} 
 based on a simple  random coding argument,
 and subsequently developed by
 Hayashi \cite{hayashi-exp} 
 based on a universal hashing argument
to establish 
the  cryptographic implication
of  channel resolvability (see, also Hayashi \cite{hayashi-wire}).
 %
 \QED
}
\echui
\bchui\label{chui:resolva1}
{\rm
We define the rates $R_B=\frac{1}{n}\log M_n$ and  $R_E=\frac{1}{n}\log L_n$, 
which is called the {\em coding rate }for Bob and the {\em resolvability rate} against Eve, respectively.
Rate $R_B$ is quite popular in channel coding, whereas rate  $R_E$, roughly speaking,  indicates
the rate of a large dice with $L_n$ faces to provide randomness  needed to 
implement an efficient {\em stochastic} encoder $\varphi_n$ to deceive Eve. 
}
\QED
\echui
\bchui\label{chui:wag-x1}
{\rm
In view of (\ref{eq:wag-re1}),
 the concatenated channels 
 $W^{n+}_B(\ssy|\ssv), W^{n+}_E(\ssz|\ssv)$ as defined by (\ref{eq:conc-cost1}) and (\ref{eq:conc-cost2}) can be written as 
\beqn
W^{n+}_B(\ssy|\ssv) &=& \sum_{\ssx \in \cX^n(\Gamma)}W^{n}_B(\ssy|\ssx)P_{X^n|V^n}(\ssx|\ssv),
\label{eq:chui10}\\
W^{n+}_E(\ssz|\ssv) &=& \sum_{\ssx \in \cX^n(\Gamma)}W^{n}_E(\ssz|\ssx)P_{X^n|V^n}(\ssx|\ssv).
\label{eq:chui11}
\eeqn
}
\echui
%

%
%
 %
%
The reason why we have introduced the concatenated channel $W^{n+}(\ssy|\ssv) $
 instead of
 the non-concatenated channel $W^n(\ssy|\ssx)$ can be seen from the following theorem.
\bteiri [Tradeoff of reliability and secrecy by concatenation]\label{teiri:tradeoff}
 {\rm 
Concatenation decreases reliability for Bob and increases secrecy against Eve.
}
\eteiri
{\em Proof:}\ \ 
 The quantity $ A_n \equiv e^{-\phi (\rho |W_B^{n+}, Q)}$ in
  (\ref{eq:hana1}) is lower bounded, by  concavity of the function $f(x) =x^{\frac{1}{1+\rho}}$, as 
 \beqn
 A_n &=& \sum_{\ssy}\left(\sum_{\ssv}Q(\ssv)\left(\sum_{\ssx}P_{X^n|V^n}(\ssx|\ssv)W_B^n(\ssy|\ssx)\right)
 ^{\frac{1}{1+\rho}}\right)^{1+\rho}\\
  & \ge & \sum_{\ssy}\left(\sum_{\ssv}\sum_{\ssx}Q(\ssv)P_{X^n|V^n}(\ssx|\ssv)W_B^n(\ssy|\ssx)
 ^{\frac{1}{1+\rho}}\right)^{1+\rho}\\
 & =& \sum_{\ssy}\left(\sum_{\ssx}P(\ssx)W_B^n(\ssy|\ssx)
 ^{\frac{1}{1+\rho}}\right)^{1+\rho},
 \eeqn 
 where $P(\ssx) =\sum_{\ssv}Q(\ssv)P_{X^n|V^n}(\ssx|\ssv)$. This implies that
 concatenation decreases reliability for the channel for Bob. On the other hand, 
 the quantity $ B_n \equiv e^{-\phi (-\rho |W_E^{n+}, Q)}$ in
  (\ref{eq:hana2c2}) is upper bounded, by  convexity of the function $g(x)=x^{\frac{1}{1-\rho}}$, as
  \beqn
 B_n &=& \sum_{\ssz}\left(\sum_{\ssv}Q(\ssv)\left(\sum_{\ssx}P_{X^n|V^n}(\ssx|\ssv)W_E^n(\ssz|\ssx)\right)
 ^{\frac{1}{1-\rho}}\right)^{1-\rho}\\
  & \le & \sum_{\ssz}\left(\sum_{\ssv}\sum_{\ssx}Q(\ssv)P_{X^n|V^n}(\ssx|\ssv)W_E^n(\ssz|\ssx)
 ^{\frac{1}{1-\rho}}\right)^{1-\rho}\\
 & =& \sum_{\ssz}\left(\sum_{\ssx}P(\ssx)W_E^n(\ssz|\ssx)
 ^{\frac{1}{1-\rho}}\right)^{1-\rho},
 \eeqn 
 which implies that concatenation increases secrecy against the channel for Eve.
 Thus, we can control the tradeoff between reliability and secrecy (usually conflicting) 
 by adequate choice of an auxiliary channel 
   $P_{X^n|V^n}$ (e.g., see Fig.\ref{fig4} later for the case of stationary memoryless wiretap channels).  
  Furthermore, it should be noted that $C_n \equiv e^{-\psi (\rho |W_E^{n+}, Q)}$ in (\ref{eq:hana2c}) also
  has such a nice tradeoff property like in the above, owing to  the convexity 
  in $W^n_E(\ssz|\ssx)$. 
\QED

{\em B. Stationary memoryless  wiretap channel with cost constraint}

So far we have studied the performance of general wiretap channels with general cost constraint $\Gamma$.
Suppose now that we are given a stationary and memoryless wiretap channel $(W^n_B, W^n_E)$,
specified by $(W_B\equiv P_{Y|X}, W_E\equiv P_{Z|X})$,
with {\em additive} cost  $c: \cX\to \sR^+$.
With this important class of channels, we attempt to bring out specific useful insights on the basis of 
 Theorem \ref{teiri:han2}.  To do so,
  let us consider the case in which $V^nX^n=(V_1X_1,\cdots, V_nX_n)$ are i.i.d. variables 
  with common joint distribution 
  \beq\label{eq:iid-dist1}
  P_{XV}(x,v)  \quad ((v, x) \in \cV\times \cX),
  \eeq
  then, the probabilities of $X^n$ and $V^n$, and 
  the conditional probability of $X^n$ given $V^n$ are written as 
  \beqn
  P_{X^n}(\ssx) &=& \prod_{i=1}^nP_X(x_i),\label{eq:iid-dist1endr1}\\
  P_{V^n}(\ssv) &=& \prod_{i=1}^nP_V(v_i),\label{eq:iid-dist1endr2}\\
  P_{X^n|V^n}(\ssx|\ssv) &=&\prod_{i=1}^nP_{X|V}(x_i|v_i),\label{eq:iid-dist1endr3}
  \eeqn
respectively,  where
 \[
  \ssx=(x_1, \cdots, x_n), \quad \ssv=(v_1, \cdots, v_n).
  \]
 It should be noted here  that $V^n$ indicates a channel input 
 for $(W^{n+}_B,W^{n+}_E)$, and $X^n$ 
 indicates  a channel input for
   $(W^n_B,W^n_E)$.
  %
%
Accordingly, these specifications define a joint probability distribution $P_{VXYZ}$ 
on $\cV\times \cX\times \cY\times \cZ$.
Also, the concatenated channel in this case is written simply as 
\beqn
W^{+}_B(y|v) &=& \sum_{x \in \cX}W_B(y|x)P_{X|V}(x|v),
\label{eq:chui10M}\\
W^{+}_E(z|v) &=& \sum_{x \in \cX}W_E(z|x)P_{X|V}(x|v).
\label{eq:chui11M}
\eeqn
Then, we have one of the key results:
 \bteiri \label{teiri:hanw2}
{\rm Let $(W^n_B, W^n_E)$ be a stationary memoryless wiretap channel with additive cost $c:\cX\to \sR^+$.
Let $P_{VXYZ}$ be a joint probability distribution 
 as above, and suppose that the  constraint $\sum_{x\in \cX}P_X(x)c(x) \le \Gamma$ on $P_X$ is satisfied.
 Then, for
any  positive integers
$M_n$, $L_n$,
 there exists a pair 
$(\varphi_n, \psi_n^B$) of encoder (satisfying cost constraint $\Gamma$) and decoder such that 
%
%
\beqn 
 \epsilon_n^B
  & \le &
\frac{2}{\alpha_n^{1+\rho}\beta_n}(M_nL_n)^{\rho}\nonumber\\
 & & \cdot \left[\sum_{y\in \cY}\left(
  \sum_{v\in \cV}q(v)\left[\sum_{x \in \cX}W_B(y|x)
P_{X|V}(x|v)e^{(1+\rho)r[\Gamma -c(x)]}\right]^{\frac{1}{1+\rho}}\right)^{1+\rho}\right]^n 
\nonumber\\
& & \label{eq:istan1}
  \eeqn
and
\beqn\label{eq:delta-wing}
\delta_n^E
 & \le & 
\frac{2}{\alpha_n^{1-\rho}\beta_n}\frac{1}{\rho L_n^{\rho}}\nonumber\\
 & & \cdot \left[\sum_{z\in \cZ}\left(
  \sum_{v\in \cV}q(v)\left[\sum_{x \in \cX}W_E(z|x)
P_{X|V}(x|v)e^{(1-\rho)r[\Gamma -c(x)]}\right]^{\frac{1}{1-\rho}}\right)^{1-\rho}\right]^n,
 \nonumber\\
& & \label{eq:istan2}
\eeqn
where we have put $q =P_V$ for simplicity, and 
$0\le \alpha_n, \beta_n \le 1$ are the constants such that 
$\lin\alpha_n \ge \lim_{n\to\infty}\beta_n = 1$ or  $1-1/\sqrt{2}$
to be specified in the proof.
}\QED
\eteiri
{\em Proof:} See Appendix \ref{appB}.
\bchui[Two secrecy functions]\label{chui:saigo2}
{\rm
So far, we have established evaluation of upper bounds (\ref{eq:hana1}) and 
(\ref{eq:hana2c2}) 
when the channel
$(W^n_B, W^n_E)$ is stationary and memoryless under cost constraint.
It should be noted, however, that we did not evaluate upper bound
 (\ref{eq:hana2c}). This is because
(\ref{eq:hana2c}) contains the term $W^n_Q(\ssz)$ with negative power $-\rho$, 
and hence upper bounding 
for  (\ref{eq:hana2c}) does not work. Thus, we prefer bound (\ref{eq:hana2c2}) rather than
bound (\ref{eq:hana2c}).
\QED
}
\echui
%
%

%

%
\bchui\label{chui:saigo3}
{\rm
%
Instead of  upper bound (\ref{eq:smooth1}) 
(in the proof of Theorem \ref{teiri:hanw2}) on the characteristic function $\chi(\ssx)$, i.e., the upper bound
\beq\label{eq:wagv1wag1}
 \chi(\ssx) \le \exp\left[(1+\rho)r\left(n\Gamma -\sum_{i=1}^nc(x_i)\right)\right],
\eeq
Gallager \cite{gall} used  
the upper bound
\beq\label{eq:smooth1m}
  \chi(\ssx) \le \exp\left[(1+\rho)r\left(\sum_{i=1}^nc(x_i) -n\Gamma+\delta\right)\right],
  \eeq
  where $\delta >0$ is an arbitrary small constant. 
  Wyner \cite{wyner1} also used  upper bound (\ref{eq:smooth1m}) for Poisson channels.  
  However, we prefer upper bound (\ref{eq:wagv1wag1}) 
   in this paper (except for in Theorems \ref{teiri:gauss-expo1} and \ref{teiri;chol-T} 
   later in Section \ref{intro-gaussian-re-se4}), 
  because it provides us with reasonable evaluation of the reliability and secrecy functions 
  for  binary symmetric wiretap channels,  for Poisson wiretap channels and also for Gaussian wiretap channels
    to be treated in this section 
  and in Sections \ref{intro-poisson-re-se2},  \ref{intro-poisson-re-conc1}
  and   \ref{intro-gaussian-re-se4}.
\QED
}
\echui

Let us now give more compact forms to (\ref{eq:istan1}) and  (\ref{eq:istan2}). 
To do so,  let us define
  a {\em reliability exponent function} (or simply, reliability function)
 $F_c(q, R_B, R_E, n)$ for Bob, and
a {\em secrecy exponent function} (or simply, secrecy function)
  $H_c(q, \rho, R_E, n)$ against Eve, as  
  \footnote{In the theory of channel coding it is the tradition to use the terminology ``reliability functionn" 
  to denote the ``optimal" one.  Therefore, more exactly, it might be recommended to use the term such as 
  ``achievable reliability exponent (function)" and  ``achievable secrecy exponent (function),"
  because  here we lack the {\em converse} results.
  However, in this paper, simply for convenience with some abuse of the notation,  
  we do not stick to the optimality and prefer to use  their shorthands,
  because in most cases
  the optimal computable formula is not known. 
  Then, the term ``{\em optimal} reliability function" with the converse makes sense.
  Similarly for the ``secrecy function."}
\beqn
\lefteqn{F_c(q, R_B, R_E,n)}\nonumber\\
&\equiv & \sup_{r\ge 0}\sup_{0\le \rho \le 1}\left(\phi (\rho |W_B, q, r)
- \rho (R_B+R_E)+\frac{\log (\alpha_n\beta_n^{1+\rho})-\rho\log 3}{n}\right),\nonumber\\
& & \label{eq:func11}
\eeqn
\beqn
\lefteqn{H_c(q, R_E,n) }\nonumber\\
& \equiv &\sup_{r\ge 0} \sup_{0<\rho <1}\left(\phi (-\rho |W_E, q, r)
+ \rho R_E+\frac{\log (\alpha_n\beta_n^{1-\rho}) +\log\rho}{n}\right),\nonumber\\
& & \label{eq:func12}
\eeqn
%
%
%
where for fixed rates $R_B, R_E$ we have set $M_n = e^{nR_B}, L_n = e^{nR_E}$, and
\beqn
\lefteqn{\phi(\rho|W_B, q, r)}\nonumber\\
 &=& -\log\left[\sum_{y\in \cY}\left(
  \sum_{v\in \cV}q(v)\left[\sum_{x \in \cX}W_B(y|x)
P_{X|V}(x|v)e^{(1+\rho)r[\Gamma -c(x)]}\right]^{\frac{1}{1+\rho}}\right)^{1+\rho}\right],
\nonumber\\
& & \label{eq:han81}
\eeqn
\beqn
\lefteqn{\phi(-\rho|W_E, q, r)}\nonumber\\
 &=& -\log\left[\sum_{z\in \cZ}\left(
  \sum_{v\in \cV}q(v)\left[\sum_{x \in \cX}W_E(z|x)
P_{X|V}(x|v)e^{(1-\rho)r[\Gamma -c(x)]}\right]^{\frac{1}{1-\rho}}\right)^{1-\rho}\right].
\nonumber\\
& & \label{eq:han82}
\eeqn
%
%
Thus, we have 
\bteiri\label{teiri:nict11}
{\rm 
Let $(W^n_B, W^n_E)$ be a stationary  memoryless wiretap channel with additive cost constraint $\Gamma$,  
 then there exists a pair 
$(\varphi_n, \psi_n^B$) of encoder (satisfying cost constraint $\Gamma$) and decoder such that 
\beqn
 \epsilon_n^B &\le & 2e^{-nF_c(q, R_B, R_E,n)},\label{eq:kanna2}\\
\delta_n^E &\le & 2e^{-nH_c(q,  R_E,n)}.\label{eq:kanna3}
\eeqn
where it is assumed that $P_X$ satisfies $\sum_{x\in \cX}P_X(x)c(x) \le \Gamma$. 
%
\QED
}
\eteiri
\bchui[Reliability and secrecy functions]\label{chui:func5}
{\rm 
The function $F_c(q, R_B, R_E,n)$ quantifies {\em performance of channel coding} 
(called the {\em random coding exponent} of  
 Gallager \cite{gall}), 
whereas the function
 $H_c(q, R_E, n)$ quantifies   {\em performance of channel resolvability}
(cf. Han and Verd\'u \cite{ver-han}, Han \cite{han-spec}, Hayashi \cite{hayashi-exp, hayashi-wire}).
}
\echui
\bchui\label{chui:neg1}
{\rm
It should be noted that, the third term in $F_c(q, R_B, R_E,n)$  on the right-hand side of (\ref{eq:func11}) and 
the third term in $H_c(q, R_E,n)$ on the right-hand of (\ref{eq:func12}) is both of the order  
$O(\frac{1}{n})$, which 
approach zero as $n$ tends to $\infty$, so that these terms do not affect the exponents. 
Actually, the term $\frac{\rho\log3}{n}$ on the right-hand side of (\ref{eq:func11}) is {\em not } needed here 
but is  needed in $F_c(q, R_B, R_E,n)$  on the right-hand side of (\ref{eq:hana51}) 
to follow under the {\em maximum criterion}.
\QED
}
\echui

%
\bchui[Non-concatenation]\label{chui:vxeq}
{\rm 
It is sometimes useful to consider the special case with $V\equiv X$ as random variables over $\cV=\cX$.
In this case the above quantities $\phi(\rho|W_B, q, r),$ $\phi(-\rho|W_E, q, r)$ ($q=P_X$)
reduce to 
\beqn
\phi(\rho|W_B, q, r) & = & 
-\log\left[\sum_{y\in \cY}\left(
  \sum_{x\in \cX}q(x)W_B(y|x)^{\frac{1}{1+\rho}}
e^{r[\Gamma-c(x)]}\right)^{1+\rho}\right],\nonumber\\
& & \label{eq:finder3}\\
\phi(-\rho|W_E, q, r) & = & 
-\log\left[\sum_{z\in \cZ}\left(
  \sum_{x\in \cX}q(x)W_E(z|x)^{\frac{1}{1-\rho}}
e^{r[\Gamma-c(x)]}\right)^{1-\rho}\right],\nonumber\\
& & \label{eq;halimeq1}
\eeqn
where the reliability function with  (\ref{eq:finder3})
with $c(x) -\Gamma$ 
 instead of $\Gamma-c(x)$
 is earlier  found in Gallager \cite{gall} and (\ref{eq:finder3}) 
 with $c(x) -\Gamma$ instead of $\Gamma-c(x)$ applied to Poisson channels
 is found in Wyner \cite{wyner1}, while 
the secrecy function with (\ref{eq;halimeq1}) intervenes
 for the first time in this paper.
 \QED
}
\echui     
%
%
Recall that, so far,  upper bounds on the error probability $\epsilon_n^B$ and the divergence distance 
$\delta_n^E$ 
 are based on the averaged criteria as mentioned in Section \ref{introduction1}.C. 
Alternatively, instead of the averaged criteria 
$\epsilon_n^B$ and $\delta_n^E$, we can define the maximum criteria 
$\mbox{{\scriptsize m}-}\epsilon_n^B$ and $\mbox{{\scriptsize m}-}\delta_n^E$ as follows.
\beqn
\mbox{{\scriptsize m}-}\epsilon_n^B &\equiv & \max_{i\in \cM_n}\Pr\{\psi_n^{B}(\varphi_n^B(i))
\neq i\},\label{eq:max-error}\\
\mbox{{\scriptsize m}-}\delta_n^E & \equiv & \max_{i\in \cM_n}D(P^{(i)}_n||\pi_n).\label{eq:max-div}
\eeqn
With these criteria, 
using Markov inequality 
\footnote{Set $\epsilon_n^B(i) = \Pr\{\psi_n^{B}(\varphi_n^B(i)),$ $\delta_n^E(i)=
D(P^{(i)}_n||\pi_n),$ then Markov inequality tells that  $\#\{i|\epsilon_n^B(i)\le 3\epsilon_n^B\}\ge 2M_n/3$
and $\#\{i|\delta_n^E(i)\le 3\delta_n^E\ge 2M_n/3.$
Therefore, $\# S_n\ge M_n/3$, where $S_n =\{i|\epsilon_n^B(i)\le 3\epsilon_n^B \mbox{\ and } \delta_n^E(i)\le 3\delta_n^E\}.$
We then keep the message set $S_n$ and throw out the rest to obtain 
Theorem \ref{teiri:max2}.
This causes the term $\frac{\rho\log 3}{n}$ to intervene on the right-hand side of (\ref{eq:func11}).}
 applied to (\ref{eq:kanna2}) and (\ref{eq:kanna3}), we obtain, instead of Theorem \ref{teiri:nict11},
 \bteiri\label{teiri:max2}
 {\rm
 Let $(W^n_B, W^n_E)$ be a stationary  memoryless wiretap channel with additive cost constraint $\Gamma$,  
 then there exists a pair 
$(\varphi_n, \psi_n^B$) of encoder (satisfying cost constraint $\Gamma$) and decoder such that 
 \beqn
 \mbox{{\scriptsize m}-}\epsilon_n^B  &\le & 6e^{-nF_c(q, R_B, R_E,n)},\label{eq:hana51}\\
\mbox{{\scriptsize m}-}\delta_n^E & \le & 6e^{-nH_c(q,  R_E, n)}\label{eq:hana61},
 \eeqn
 where it is assumed that $P_X$ satisfies $\sum_{x\in \cX}P_X(x)c(x) \le \Gamma$.
 }\QED
 \eteiri
 \bchui[Average vs. maximum criteria]\label{chui:max1}
 {\rm
 Bound  (\ref{eq:hana51}) is well known
  in channel coding (cf. Gallager \cite{gall}), whereas 
  bound (\ref{eq:hana61})  is taken
   into consideration for the first time in this paper. \\
   \ \ 
   In channel coding,  which of the averaged $\epsilon_n^B$
    or the maximum $\mbox{{\scriptsize m}-}\epsilon_n^B$ we should  take would 
    be rather a matter of preference
    or the context.
  On the other hand, however, which of 
  the averaged $\delta_n^E$
    or the maximum $\mbox{{\scriptsize m}-}\delta_n^E$ we should take is 
     a serious matter from the viewpoint of secrecy. This is because, even with small $\delta_n^E$,
      we cannot exclude
     a possibility that  the divergence distance $D(P^{(i)}_n||\pi_n)$ is very large for some particular 
      $i\in \cM_n,$ and hence 
     $\mbox{{\scriptsize m}-}\delta_n^E$ is also very large, which implies that the message $i$ 
     is not saved from a serious risk of successful decryption by Eve. On the other hand, with small 
     $\mbox{{\scriptsize m}-}\delta_n^E$, every message $i\in\cM_n$ is guaranteed to be kept
     highly confidential against Eve as well.
     Thus, we prefer the criterion $\mbox{{\scriptsize m}-}\delta_n^E$
     as well as $\mbox{{\scriptsize m}-}\epsilon_n^B$  in this paper.
      \QED
     }\echui
    

%
%
%
In view of Remark \ref{chui:neg1}, we are tempted to go further  over the properties of  the functions
$F_c(q,  R_B, R_E,n),$ $ H_c(q,  R_E, n)$. In particular, we are interested in the behavior of
the functions
$F_c(q, R_B, R_E, +\infty)$ and $H_c(q,  R_E, +\infty).$ In this connection, we have following lemma,
where we let $I(q,W)$ denote the mutual information between the input $q$ and   
its output via the channel  $W$.
\bhodai\label{hodai:wag1q}
{\rm \mbox{} Assume that $\sum_{x\in \cX}P_X(x)c(x)\le \Gamma$ and $I(q, W_B^+)>0$, then
\begin{enumerate}
\item
$F_c(q, R_B, R_E,+\infty)=0 \mbox{\ at\ } R_B+R_E =I(q, W_B^+); $ 
\item
$H_c(q, R_E, +\infty) = 0 \mbox{\ at\ } R_E=I(q, W^+_E);$
\item
$F_c(q, R_B, R_E,+\infty)$ is a monotone strictly decreasing  {\em positive} convex  function of $R_B+R_E$ 
for $R_B+R_E< I(q, W^+_B),$ and $F_c(q, R_B, R_E,+\infty)=0$ for $R_B+R_E\ge I(q, W^+_B);$
\item
$H_c(q, R_E, +\infty)$ is a monotone strictly increasing {\em positive} convex  function of $R_E$ 
for $R_E > I(q, W^+_E),$
and $H_c(q, R_E, +\infty)=0$ for $R_E \le I(q, W^+_E).$ 
\end{enumerate}
}
\ehodai

{\em Proof:}  See Appendix \ref{appC}.   This lemma is used later to prove 
Theorems \ref{teiri:wagcont1}, \ref{teiri:wagcont41}. \QED

{\em C.  Strengthening of Theorem \ref{teiri:hanw2}}

%

Let us now consider strengthening Theorem \ref{teiri:hanw2}.
Since it holds that
\beq\label{eq:tuikac1}
\sum_{x\in \cX}P_X(x)c(x) = \sum_{v\in \cV}P_V(v)\overline{c}(v),
\eeq
where
\beq
\overline{c}(v) = \sum_{x\in \cX}c(x)P_{X|V}(x|v), 
\eeq
we see that  $\Exp [c(X)] = \Exp [\overline{c}(V)]$, and hence   $\sum_{x\in \cX}P_X(x)c(x) \le \Gamma$
is equivalent to $\sum_{v\in \cV}P_V(v)\overline{c}(v) \le \Gamma$.
Therefore,  it is concluded again by virtue of
 the central limit theorem that, 
 as in 
 the proof (Appendix \ref{appB}) of Theorem \ref{teiri:hanw2}, 
  we have
\[
  \lim_{n\to\infty}\overline{\mu}_n =1\mbox{\quad  with\ }
 \overline{\mu}_n= \sum_{\ssv}\overline{\chi}(\ssv)\prod_{i=1}^nP_{V}(v_i),
 \]
 where  $\ssv=(v_1, v_2,\cdots, v_n) \in \cV^n$ and, with any constant $a$ such that $1/2<a<1$
  \beq\label{eq:ransuu1E}
  \overline{\chi}(\ssv) = \left\{
  \begin{array}{cl}
  1 & \mbox{for} \  \sum_{i=1}^n \overline{c}(v_i) \le n\Gamma +n^a,\\
  0 & \mbox{otherwise},
  \end{array}
  \right.
  \eeq
  so that $\cT_0$ in the proof (Appendix \ref{appB}) of Theorem \ref{teiri:hanw2}  can be replaced by 
  $\cT_1\equiv \cT_0\cap \{\ssv\in \cV^n|\overline{\chi}(\ssv)=1\}$
  without affecting the process of the proof.
  %
  This observation means that cost constraint $\Gamma$ (with cost $c(x)$) on  $P_{X^n}$ 
of the concatenated channel
  $(W^{n+}_B, W^{n+}_E)$ is consistent with  cost constraint $\Gamma$ 
(with cost $\overline{c}(v)$) on  $P_{V^n}$ 
of  the concatenated channel $(W^{n+}_B, W^{n+}_E)$.
 Thus, by introducing the upper bound
 %
   \beq\label{eq:smooth1h}
  \overline{\chi}(\ssv) \le \exp\left[s\left(n\Gamma-\sum_{i=1}^n\overline{c}(v_i)+n^a \right)\right],
  \eeq
  where $s\ge 0$ is an arbitrary number,
 we can strengthen upper bounds (\ref{eq:istan1}) and (\ref{eq:istan2}) as:
\bteiri\label{chui:tuika3}
{\rm With the same notation and assumption as in Theorem \ref{teiri:hanw2}, we have
\beqn 
  \lefteqn{\epsilon_n^B}\nonumber\\
   & \le &
\frac{2e^{s(1+\rho)n^a}}{\alpha_n^{1+\rho}\beta_n}(M_nL_n)^{\rho}\nonumber\\
 & & \cdot \left[\sum_{y\in \cY}\left(
  \sum_{v\in \cV}q(v)e^{s[\Gamma -\overline{c}(v)]}\left[\sum_{x \in \cX}W_B(y|x)
P_{X|V}(x|v)e^{(1+\rho)r[\Gamma -c(x)]}\right]^{\frac{1}{1+\rho}}\right)^{1+\rho}\right]^n, 
\nonumber\\
& & \label{eq:istan1x}
  \eeqn
 \beqn\label{eq:delta-wing}
\lefteqn{\delta_n^E}\nonumber\\
 & \le & 
\frac{2e^{s(1-\rho)n^a}}{\alpha_n^{1-\rho}\beta_n}\frac{1}{\rho L_n^{\rho}}\nonumber\\
 & & \cdot \left[\sum_{z\in \cZ}\left(
  \sum_{v\in \cV}q(v)e^{s[\Gamma -\overline{c}(v)]}\left[\sum_{x \in \cX}W_E(z|x)
P_{X|V}(x|v)e^{(1-\rho)r[\Gamma -c(x)]}\right]^{\frac{1}{1-\rho}}\right)^{1-\rho}\right]^n.
 \nonumber\\
& & \label{eq:istan2x}
\eeqn
}
\eteiri
\bchui\label{chui:saigou-wadl}
{\rm 
Notice here that the terms $e^{s(1+\rho)n^a}$ and $e^{s(1-\rho)n^a}$ 
in (\ref{eq:istan1x}) and (\ref{eq:istan2x}) do not affect
the exponents of exponential decay in $n$ for $\epsilon_n^B$ and $\delta_n^B$.
%
%
Accordingly, instead of (\ref{eq:func11}), (\ref{eq:func12}) and (\ref{eq:han81}), (\ref{eq:han82}), let us define 
\beqn
\lefteqn{F_c(q, R_B, R_E,n)}\nonumber\\
&\equiv &\sup_{s\ge0, r\ge 0} \sup_{0\le \rho \le 1}\left(\phi (\rho |W_B, q, s, r)
- \rho (R_B+R_E)+\frac{\log (\alpha_n\beta_n^{1+\rho})-\rho\log 3 -sn^a}{n}\right),\nonumber\\
& & \label{eq:func11-f}
\eeqn
\beqn
\lefteqn{H_c(q, R_E,n) }\nonumber\\
& \equiv &\sup_{s\ge 0, r\ge 0} \sup_{0<\rho <1}\left(\phi (-\rho |W_E, q, s, r)
+ \rho R_E+\frac{\log (\alpha_n\beta_n^{1-\rho}) +\log\rho - sn^a}{n}\right),\nonumber\\
& & \label{eq:func12-f}
\eeqn
\beqn
\lefteqn{\phi(\rho|W_B, q, s, r)}\nonumber\\
 &=& -\log\left[\sum_{y\in \cY}\left(
  \sum_{v\in \cV}q(v)e^{s[\Gamma -\overline{c}(v)]}\left[\sum_{x \in \cX}W_B(y|x)
P_{X|V}(x|v)e^{(1+\rho)r[\Gamma -c(x)]}\right]^{\frac{1}{1+\rho}}\right)^{1+\rho}\right],
\nonumber\\
& & \label{eq:han81x}
\eeqn
\beqn
\lefteqn{\phi(-\rho|W_E, q, s, r)}\nonumber\\
 &=& -\log\left[\sum_{z\in \cZ}\left(
  \sum_{v\in \cV}q(v)e^{s[\Gamma -\overline{c}(v)]}\left[\sum_{x \in \cX}W_E(z|x)
P_{X|V}(x|v)e^{(1-\rho)r[\Gamma -c(x)]}\right]^{\frac{1}{1-\rho}}\right)^{1-\rho}\right].
\nonumber\\
& & \label{eq:han82x}
\eeqn
Then,  Theorems \ref{teiri:nict11},  \ref{teiri:max2} with the $F_c(q, R_B, R_E,n)$, $H_c(q, R_E,n)$ thus modified
are guaranteed to 
give the performance  better  than or equal to the original version only
with the term $e^{(1+\rho)r[\Gamma -c(x)]}$.
\   However, here we do not go into the details of its analysis.
The case with $r=0$  will be used later in Section \ref{intro-poisson-re-conc1} to establish the 
reliability and secrecy functions for concatenated Poisson wiretap channels.
}
\echui
%
 %
 %
%
 %
%
%

   %
   %
   %
   
 {\em  D. $\delta$-secrecy capacity with cost constraint}  
   
  Suppose that we are given a stationary memoryless wiretap channel $(W_B, W_E)$,
  and consider any Markov chain such that
  \beq\label{eq:wagcot2}
  V\to X\to YZ, \quad P_{Y|X} = W_B, \ P_{Z|X} = W_E.
  \eeq 
 %
   Then, we have 
   \bteiri\label{teiri:wagcont1}
   {\rm
   Let $(W_B, W_E)$ be a stationary memoryless wiretap channel with cost constraint $\Gamma$. Then, the
   $\delta$-secrecy capacity (cf. Section \ref{intro-geri1}.C) is given by
   \beq\label{eq:wagnerp2}
   \delta\mbox{-}C_s(\Gamma) = \sup_{VX:\Exp[c(X)]\le \Gamma}(I(V; Y) - I(V; Z))
   \eeq
   under the maximum criterion $(\mbox{m-}\epsilon_n^B, \mbox{m-}\delta_n^E), $
   where  the supremum on the right-hand side ranges over all $VX$ satisfying (\ref{eq:wagcot2}) and 
   $\Exp[c(X)]\le \Gamma$. \footnote{
   After the submission of this paper, Hou and Kramer \cite{hou-kra}  independently obtained formula (\ref{eq:wagnerp2})
   for the case  {\em without} cost constraint (i.e., $c(x)=\Gamma=1$ for all $x\in \cX$)
   under the {\em finite} alphabet assumption; they call it the {\em effective secrecy capacity}. 
   }

   }
    \eteiri
   
  {\em Proof:} \mbox{}
  It is not difficult to see that the converse part
  \beq\label{eq:wagnerp3}
  \delta\mbox{-}C_s(\Gamma) \le \sup_{VX:\Exp[c(X)]\le \Gamma}(I(V; Y) - I(V; Z))
   \eeq
   holds (cf. \cite{csis-kor-2nd}, \cite{bloch}).
Therefore, it suffices only to show the opposite inequality (achievability part). 
     To do so,
     let 
      $V_0 \to X_0 \to Y_0Z_0$
      denote the Markov chain to attain the supremum 
     on the right-hand side of (\ref{eq:wagnerp2}) and  let $(W_B^+, W_E^+)$  indicate 
     the concatenated wiretap channel of
    $(W_B, W_E)$ using the auxiliary channel $P_{X_0|V_0}$.
    Then, with $q=P_{V_0}$  it is easy to  observe that $I(V_0; Y_0)= I(q, W_B^+)$ and  $I(V_0; Z_0)= I(q, W_E^+)$.
    Furthermore, with an arbitrarily small number $\tau >0$ we set as 
    $R_B + R_E = I(q, W_B^+) -\tau$ and $R_E = I(q, W_E^+) +\tau$, and hence
    $R_B=I(q, W_B^+) -I(q, W_E^+) -2\tau$.  
    With these rates $R_B, R_E$ Lemma \ref{hodai:wag1q} guarantees that 
    \[
    F_c(q, R_B, R_E,+\infty)>0,\quad H_c(q, R_E, +\infty)>0,
     \]
     which together with
    Theorem \ref{teiri:max2} concludes that both of the error probability $\epsilon_n^B$ and the divergence distance
    $\delta_n^E$ exponentially decay with increasing $n$, provided that $n$ is sufficiently large.
    Thus, the rate  $R_B=I(q, W_B^+) -I(q, W_E^+) -2\tau$ is $\delta$-achievable, that is, 
    $R_B=I(V_0; Y_0) -I(V_0; Z_0)-2\tau$ is $\delta$-achievable 
     under the maximum criterion $(\mbox{m-}\epsilon_n^B, \mbox{m-}\delta_n^E)$ (cf. Theorem \ref{teiri:max2}).
     %
%
 \QED

    Now we are ready to go to the problem  of the secrecy capacity when the wiretap channel $(W_B, W_E)$ is 
     more capable:
     
    \bteigi\label{teigi:wagner4} 
    {\rm
    Let $(W_B, W_E)$ be a stationary memoryless wiretap channel. If $I(X;Y)\ge I(X;Z)$ 
    holds for any input variable $X$, we say that the wiretap channel is {\em more capable}.
    %
     }\QED
     \eteigi
   \bteiri\label{teiri:wagcont41}
   {\rm
   If a stationary memoryless wiretap channel $(W_B, W_E)$ with cost constraint $\Gamma$ is  
   more capable, then the
   $\delta$-secrecy capacity (cf. Section \ref{intro-geri1}.C) is given by
   \beq\label{eq:wagnerp2c}
   \delta\mbox{-}C_s(\Gamma) = \sup_{X:\Exp[c(X)]\le \Gamma}(I(X; Y) - I(X; Z))
   \eeq
   under the maximum criterion $(\mbox{m-}\epsilon_n^B, \mbox{m-}\delta_n^E), $
   where  the supremum on the right-hand side ranges over all $X$ satisfying (\ref{eq:wagcot2}) and 
   $\Exp[c(X)]\le \Gamma$.
   }
   \eteiri   

 {\em Proof:}\ \ 
  In the light of Theorem \ref{teiri:wagcont1}, it suffices to show that
   \[
   I(V; Y) - I(V; Z) \le  I(X;Y)- I(X;Z),
   \]
  which is seen as follows. 
  \beqn
   I(V;Y)- I(V;Z) &=& I(VX;Y)-I(X;Y|V) - I(VX;Z)+I(X;Z|V)\nonumber\\
    &=& I(X;Y) -I(X;Z) -(I(X;Y|V)-I(X;Z|V))\nonumber\\
    &=& I(X;Y) -I(X;Z) -\sum_{v\in\cV}P_V(v)(I(X;Y|V=v)-I(X;Z|V=v))\nonumber\\
   &\le&I(X;Y) -I(X;Z),
   \eeqn
where in the last step we have used the more capability.
\QED  
   %
   %
%
%
 %
%
%
%
 %
 \section{Tradeoff of reliability and secrecy}\label{sec:tradeoff}  
\begin{figure}[htbp]
\begin{center}
\includegraphics[width=70mm]{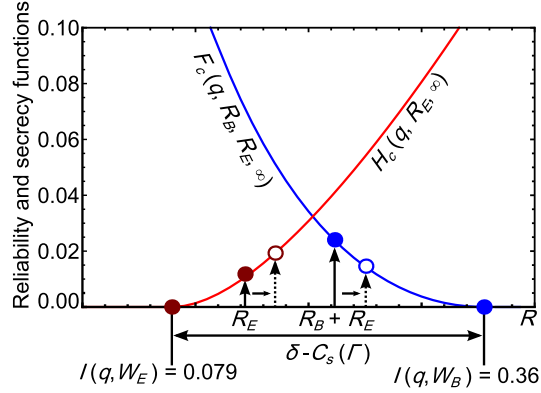}
\end{center}
\caption{Reliability and secrecy functions with cost constraint for non-concatenated BSC  and rate shifting
         ($\vep_y = 0.1$, $\vep_z = 0.3$, $c(0) = 1$, $c(1) = 2$, $\Gamma = 1.4, q(1)=0.4$).}
\label{fig5}
\end{figure}

Thus far, we have established the general computable formulas for the reliability function 
$F_c(q,R_B, R_E, +\infty)$ and the secrecy function 
$H_c(q, R_E, +\infty)$ with the stationary memoryless wiretap channel under cost constraint.
From the viewpoint of secure communications, 
these  should be regarded as a {\em pair} of functions but not as separate ones, 
which then enables us to {\em quantify} the tradeoff of reliability and secrecy.
It should be emphasized that in wiretap channel coding it is one of the crucial problems
how to control tradeoff of reliability and secrecy.
In order to elucidate this specifically, in this section we 
 focus on  wiretap channels $(W_B, W_E)$ consisting of two BSC's (Binary Symmetric  Channel)
 with crossover probabilities $\vep_y$ for Bob and $\vep_z$ against Eve ($0\le \vep_y < \vep_z\le 1/2$), 
 because this class of wiretap channels are quite tractable but still very informative.\\
%
%
 %
%
%
%
\ \ 
On the basis of the paired functions, we can consider several ways to control the tradeoff
of reliability and secrecy. Typical four ways are considered and discussed in the following.
A  typical pair of reliability and secrecy  functions in this BSC case is depicted in Fig.\ref{fig5}. 
It should be noted here that 
for any pair of BSC's one is  degraded (and hence also is more capable) with respect to  the other one, 
so that in calculating the $\delta$-secrecy capacity $\delta$-$C_s(\Gamma)$
we can invoke formula (\ref{eq:wagnerp2c})  with $q=P_X$
in Theorem \ref{teiri:wagcont41} (along with Lemma \ref{hodai:wag1q}).  More specifically, 
let $q$ indicate the input  maximizing
$I(q, W_B)-I(q, W_E)$ (while satisfying the condition $\sum_{x}q(x)c(x)\le \Gamma$),
 then this  $I(q, W_B)-I(q, W_E)$ gives the $\delta$-secrecy capacity
$\delta$-$C_s(\Gamma)$, as is depicted in Fig.\ref{fig5}. The input $q$  in  all the figures to follow
denotes the maximizing one in this sense.

  {\em A. Tradeoff of reliability and secrecy by rate shifting}

 First of all, Fig.\ref{fig5} immediately suggests a primitive and simple way ({\em rate shifting})
 of the tradeoff: moving  $R_E$ 
 (resolvability rate)
 while keeping $R_B$ (coding rate) {\em unchanged}
 enables us to control the tradeoff between the reliability exponent and the secrecy exponent, i.e., increasing $R_E$
 causes stronger secrecy but with lower reliability, whereas decreasing $R_E$ causes higher reliability but with weaker secrecy.
A technological intuition is that increasing secrecy requires ``expanding" each signaling point into multiple, 
 which is harmful from a reliability standpoint.

%
%
      

{\em B. Tradeoff of reliability and secrecy by rate exchange}

\begin{figure}[htbp]
\begin{center}
\includegraphics[width=70mm]{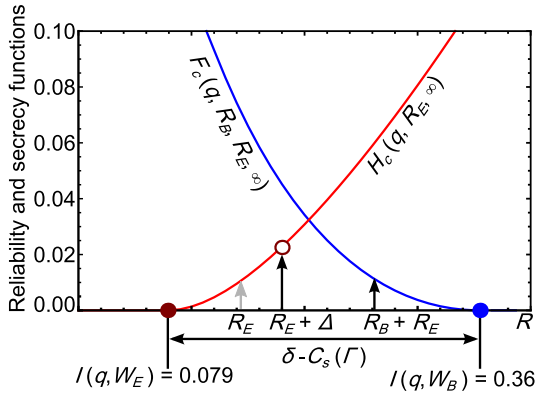}
\end{center}
\caption{Tradeoff by rate exchange: let $R_E \to R_E + \Delta$
         and $R_B \to R_B - \Delta$ ($R_B+R_E$ remains unchanged), 
         then secrecy against Eve increases by $\Delta$ and 
         rate $R_B$ decreases by $\Delta$ but at no expense of reliability for Bob;
         $\vep_y=0.1, \vep_z=0.3, c(0)=1, c(2)=2, \Gamma=0.4, q(1)=0.4$}
\label{exchange}
\end{figure}

One more way to 
control such a tradeoff is to handle rates $R_B, R_E$, where 
the enhancement of secrecy is attained at the expense of rate $R_B$ but not at the expense of reliability: 
with the same exponents $F_c(q, R_B, R_E,+\infty),$ $H_c(q, R_E, +\infty)$ as above,
we let $R_E$ increase while keeping the sum $R_B+R_E$ {\em unchanged},
which implies decrease of rate $R_B$ but no expense of reliability, because then
the value of $H_c(q, R_E, +\infty)$  increase but that of $F_c(q, R_B, R_E,+\infty)$
remains {\em unchanged}.  See Fig.\ref{exchange}.
A technological meaning of this tradeoff is as follows: suppose that a codeword
consists of $R_B$ information bits, $R_E$ random bits  and $R_H$ check bits in a memory device.
The operation of rate exchange corresponds to  shifting of the partition between
$R_B$ information bits and $R_E$ random bits, while $R_B +R_E$ is {\em unchanged}.

%

%

%

%
 {\em C. Tradeoff of reliability and secrecy by concatenation}

\begin{figure}[htbp]
\begin{center}
\includegraphics[width=70mm]{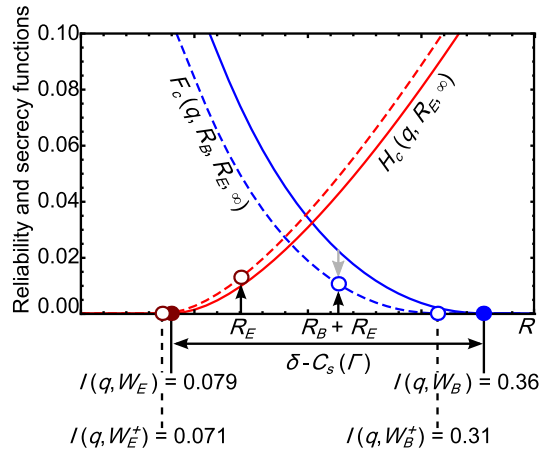}
\end{center}
\caption{Tradeoff by concatenation:  reliability and secrecy functions for non-concatenated (solid lines) 
         and concatenated (dashed lines); $\vep_v$ = 0.025, $\vep_y = 0.1$, $\vep_z = 0.3$, 
         $c(0)=1, c(1)=2, \Gamma=1.4, q(1)=0.4$
         where
          reliability for Bob decreases but secrecy against Eve increases with fixed $R_B, R_E$.}
\label{fig4}
\end{figure}

%
%
%
 Now,  let us consider another BSC with crossover probability $\vep_v$
as an auxiliary channel $P_{X|V}:\cV\to \cX$ for concatenation. 
Then, the reliability and secrecy functions for both  of  the non-concatenated and  
concatenated BSC wiretap channels
can be depicted together  in Fig.\ref{fig4}.  
We observe from this figure that, with {\em fixed} rates $R_B, R_E$, 
concatenation makes reliability for Bob decrease but
makes secrecy against Eve increase, which is guaranteed  by  Theorem \ref{teiri:tradeoff}.
Especially, we can compute numerically this tradeoff of reliability and secrecy in terms of their exponents
$F_c(q,R_B, R_E, +\infty)$ and $H_c(q, R_E, +\infty)$.
%
%
Notice, from the technological point of view, the auxiliary channel can be simulated by using
a random number generator implemented by Alice.
%
%
More
importantly, the implementation of concatenation (auxiliary channel) using
a random number generator is technologically indispensable to achieve
the secrecy capacity when the channel is {\em not} more capable (or {\em not} less noisy). So, the concatenation
technique has two kind of technological advantages, one is to
control the tradeoff and the other to achieve the secrecy capacity.
%
%
%
%
%
%
%

%
%
%
%
%
%
%
%
%

%
%
%

%
{\em D. Tradeoff of reliability and secrecy by change of cost constraint}

\begin{figure}[htbp]
\begin{center}
\includegraphics[width=70mm]{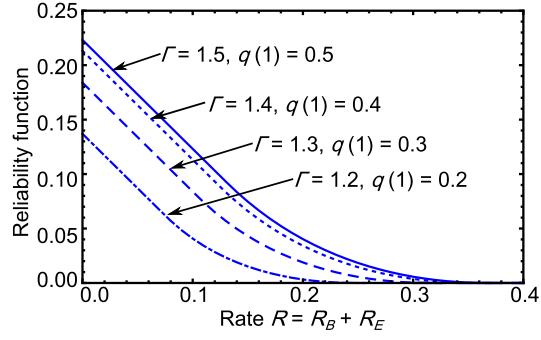}
\end{center}
\caption{Reliability function for non-concatenated BSC with varied  cost constraint $\Gamma$.
The reliability function curve moves upward as allowed cost $\Gamma$ becomes large 
($\vep_y = 0.1$, $c(0) = 1$, $c(1) = 2$).}
\label{fig-cost-ex1}
\end{figure}

\begin{figure}[htbp]
\begin{center}
\includegraphics[width=70mm]{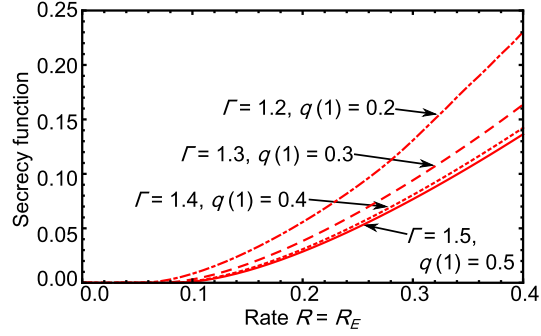}
\end{center}
\caption{Secrecy function  for non-concatenated BSC with varied  cost constraint $\Gamma$.
The secrecy function curve moves downward as allowed cost $\Gamma$ becomes large 
 ($\vep_z = 0.3$, $c(0) = 1$, $c(1) = 2$).}
\label{fig7}
\end{figure}

The fourth way to control the tradeoff between reliability and secrecy is to change cost constraint $\Gamma$.
Generally speaking, relaxing cost constraint $\Gamma$ brings about  increase of reliability and decrease of secrecy, 
whereas
strengthening cost constraint $\Gamma$ brings about  decrease of reliability and increase of secrecy, 
as is shown in Figs. \ref{fig-cost-ex1} and \ref{fig7}. This is because relaxing of cost constraint will increase 
the  ability of implementing, 
based on adaptive fitting of  the input distribution $q$ to the allowed cost $\Gamma$, good
  codes with {\em finer} decoding regions at the 
{\em fixed} rate $R_B+R_E$,
and hence leading to {\em higher reliability}
 and at the same time leading to  {\em weaker  secrecy} at the {\em fixed} rate $R_E$.
Notice here that  {\em finer} decoding regions will decrease
 the ability of deceiving Eve;  and vice versa.
From the technological point of view, this implies that {\em cheaper} cost can attain {\em stronger} secrecy
but with {\em lower} reliability.

%
%


%
%
%
\section{Secrecy capacity of Poisson wiretap channel}\label{intro-poisson}
%
%
%
In this section, we consider application of  Theorem \ref{teiri:wagcont41} 
to the Poisson wiretap channel to determine its secrecy capacity. First of all, 
let us define the Poisson wiretap channel 
(cf. \cite{wyner1}, \cite{lapidoth}, \cite{wagner}).
The input process to the Poisson channel is a waveform denoted by $X_t$ $(0\le t\le T)$ 
satisfying $X_t \ge 0$
for all $t$, where $T$ is an arbitrarily large time span. 
We assume that the input process is not only peak power limited, i.e., $0 \le X_t \le 1$ for all $t$
but also  average power limited, i.e., 
\beq\label{ew:power-const1}
\frac{1}{T}\int_0^T X_tdt \le \Gamma  \quad (0\le \Gamma \le 1).
\eeq
The output signal  to be received by the legitimate receiver Bob  is a  Poisson counting process $Y_t$
 $(0\le t\le T)$
with instantaneous rate $A_yX_t +\lambda_y$ ($\lambda_y \ge 0$ is  the dark current, and
$A_y >0$ specifies attenuation of signal)
 such that
\beq\label{eq:po1}
Y_{t=0}=0,
\eeq
and, for $0\le t, t+\tau\le T$ ($\tau >0$),
\beq\label{eq:po2}
\Pr\{Y_{t+\tau} - Y_t =j\} = \frac{e^{-\Lambda}\Lambda^j}{j!}\quad (j=0,1,2,\cdots),
\eeq
where 
\beq\label{eq:po3}
\Lambda =\int_t^{t+\tau}(A_yX_u +\lambda_y)du.
\eeq
Similarly, the output signal to be received by the eavesdropper Eve is 
a  Poisson counting process $Z_t$
 $(0\le t\le T)$
with instantaneous rate $A_zX_t +\lambda_z$.

We now want to discretize the continuous time process like this into a discrete time process
in order to make the  problem more tractable with asymptotically negligible loss of  performance.
To do so, we follow the way that Wyner \cite{wyner1} has demonstrated, 
and for the reader's convenience we review here his formulation to be exact.
Let $\Delta>0$ be an arbitrary very small  constant. Then, 
we assume the following.

{\em a})  The channel input $X_t$ is constant for $(i-1)\Delta < t \le i\Delta$ $(i=1,2,\cdots)$,
and $X_t$ takes only the values $0$ or $1$. For $i=1,2, \cdots$, define as $x_i = 0$  or $1$ 
according as $X_t=0$ or $1$ in the interval $((i-1)\Delta, i\Delta ]$.\\
{\em b})  Bob observes only the samples $Y_{i\Delta}$ $(i=1,2,\cdots)$, 
and define as
$y_i = 1$ if $Y_{i\Delta} -Y_{(i-1)\Delta}=1$; $y_i =0$ otherwise. Here, $Y_0=0.$\\
{\em c})  Eve observes only the samples $Z_{i\Delta}$ $(i=1,2,\cdots)$, 
and define as
$z_i = 1$ if $Z_{i\Delta} -Z_{(i-1)\Delta}=1$; $z_i =0$ otherwise. Here, $Z_0=0.$

Owing to the discretization under  assumptions a), b), c), we have two channels $W_B, W_E$
for Bob and Eve, respectively, i.e.,
two-input two-output stationary memoryless discrete channels  such as 
$W_B: x_i\to y_i$ and $W_E: x_i\to z_i$, whose transition probabilities are given,
up to the  order $O(\Delta)$,  as 
\beqn
W_B(1|0) &=&\lambda_y\Delta e^{-\lambda_y\Delta}\nonumber\\
&\simeq & 
 \lambda_y\Delta  =s_yA_y\Delta ,\label{eq;po5}\\
W_B(1|1) &=& (A_y+\lambda_y)\Delta e^{- (A_y+\lambda_y)\Delta} \nonumber\\
&\simeq &(A_y+\lambda_y)\Delta 
= A_y(1+s_y)\Delta ;\label{eq;po6}
\eeqn
\beqn
W_E(1|0) &=& \lambda_z\Delta e^{-\lambda_z\Delta}\nonumber\\
&\simeq & \lambda_z\Delta = s_zA_z\Delta   ,\label{eq;po7}\\
W_E(1|1) &=& (A_z+\lambda_z)\Delta e^{-(A_z+\lambda_z)\Delta}
\nonumber\\
&\simeq& (A_z+\lambda_z)\Delta
= A_z(1+s_z)\Delta ,\label{eq;po8}
\eeqn
where we have put
\beq\label{eq:post1}
s_y = \frac{\lambda_y}{A_y}, \quad s_z = \frac{\lambda_z}{A_z}.
\eeq
Furthermore,  a given fixed constant $\Delta>0$ small enough, define the whole time interval  $T=n\Delta$,
where $n$ denotes the block length of the DMC.
Then, the power constraint  (\ref{ew:power-const1}) is equivalent to
\beq\label{eq:postq1}
\nth \sum_{i=1}^nc(x_i) \le \Gamma,
\eeq
where the additive cost $c(x)$ is defined as $c(x)=x$ for $x=0, 1$.
We are now almost ready to apply Theorem \ref{teiri:wagcont41} and Theorem \ref{teiri:max2}
to find secrecy capacities and reliability/secrecy functions.

However, since Theorem \ref{teiri:wagcont41} holds only for more capable  channels, 
we need to impose some  restriction on  the class of Poisson wiretap channels as above formulated. 
In this connection, we introduce the concept of  degradedness of  channels as follows:
\bteigi[{\rm \cite{csis-kor-2nd}}]\label{teigi:degr1}
{\rm
A Poisson wiretap channel $(W_B, W_E)$ is said to be (statistically)
{\em degraded}
\footnote{More exactly, we should say that  the channel $W_E$ is  degraded 
with respect to the channel $W_B$. Here, with abuse of notation, we simply say that 
$(W_B, W_E)$ is  degraded.}
 if there exists an auxiliary channel $T:\cY\to\cZ$
such that 
\beq\label{eq:deg-cond1}
W_E(z|x) = \sum_{y\in \cY}W_B(y|x)T(z|y).
\eeq
}
\eteigi
In this connection, we have the following theorems:
\bteiri[{\rm \cite{lapidoth}, \cite{wagner}}]\label{teiri:degraded1}
{\rm 
A Poisson wiretap channel is  degraded if
\beq\label{eq:degc1}
A_y \ge A_z
\eeq
and
\beq\label{eq:degc2}
\frac{\lambda_y}{A_y} \le \frac{\lambda_z}{A_z}.
\eeq
}
\eteiri
\bteiri[{\rm \cite{csis-kor-2nd}}]\label{teiri:degraded2}
{\rm 
A Poisson wiretap channel is more capable  if it is  degraded.
\QED
}
\eteiri
Thus, in the sequel, we  confine ourselves to the class of  Poisson
wiretap channels satisfying (\ref{eq:degc1}) and (\ref{eq:degc2}) 
to guarantee the application of Theorem \ref{teiri:wagcont41}, where we assume 
that at least one of them
holds with strict inequality; otherwise the problem is trivial. 

With these preparations, we now go to the problem of determining the secrecy capacity.
Let $X$ be a channel input, and $Y, Z$ be the channel output via $W_B, W_E$, respectively, 
due to $X$. 
Following Wyner \cite{wyner1} with $q=\Pr\{X=1\}$, we directly compute the mutual informations to have
\beqn\label{eq:wyner1}
I(X;Y) 
&=& \Delta A_y[-(q+s_y)\log (q+s_y) + q(1+s_y)\log (1+s_y) \nonumber\\
& & + (1-q)s_y\log s_y] \stackrel{\Delta}{=} f(q),\label{eq: wyan8}
\eeqn
\beqn\label{eq:wyner5}
I(X;Z) 
&=&  \Delta A_z[-(q+s_z)\log (q+s_z) + q(1+s_z)\log (1+s_z) \nonumber\\
& & + (1-q)s_z\log s_z] \stackrel{\Delta}{=} g(q),\label{eq: wyan9}\\
\sigma (q) &\stackrel{\Delta}{=}& f(q)-g(q).\label{eq: wyan10}
\eeqn
Then, it is evident that
\beq\label{eq:wyn12}
\sigma(0)= \sigma (1) = 0.
\eeq
Moreover, 
\beq\label{eq:wyn13}
\sigma^{{\prime}{\prime}}(q)= -\frac{\Delta A_y}{q+s_y}+\frac{\Delta A_z}{q+s_z} < 0,
\eeq
where the inequality follows from (\ref{eq:degc1}) and (\ref{eq:degc2}).
Therefore, $\sigma (q)$ is strictly concave and takes the maximum value at the 
unique $q=q^*$ in the interval $(0,1)$
with $\sigma^{\prime}(q^*) =0$. Thus, we have  one of the main results as follows. 

%
%
%
\bteiri\label{teiri:wyn1}
{\rm
The  $\delta$-secrecy capacity with cost constraint
$\delta$-$C_s(\Gamma)$ per second of the Poisson wiretap channel $(W_B, W_E)$ 
is given by  
\beqn\label{eq:karuwaza23}
\lefteqn{\delta\mbox{-}C_s(\Gamma)}\nonumber\\
&=&
\log\frac{(q^*_{\Gamma}+s_z)^{(q^*_{\Gamma}+s_z)A_z}}
{(q^*_{\Gamma}+s_y)^{(q^*_{\Gamma}+s_y)A_y}}
+  \log\frac{s_y^{s_yA_y}}{s_z^{s_zA_z}}\nonumber\\
& & +  q^*_{\Gamma}\left(\log\frac{(q^*+s_y)^{A_y}}{(q^*+s_z)^{A_z}} + A_y - A_z\right)
\eeqn
under the maximum criterion (m-$\epsilon_n^B$, m-$\delta_n^E$),
where
$q=q^*$ is the unique solution in $(0,1)$ of the equation:
\beq\label{eq:keisan70}
\frac{(A_yq^*+\lambda_y)^{A_y}}{(A_zq^*+\lambda_z)^{A_z}}
= e^{A_z-A_y}\frac{(A_y+\lambda_y)^{A_y+\lambda_y}}{(A_z+\lambda_z)^{A_z+\lambda_z}}
\frac{\lambda_z^{\lambda_z}}{\lambda_y^{\lambda_y}},
\eeq
and 
\beq\label{eq:karuwaza9}
q^*_{\Gamma} = \min (q^*, \Gamma).
\eeq
}
\eteiri
{\em Proof:}:\ \ 
We develop $\sigma(q)$ in (\ref{eq: wyan10}) as follows:
\beqn
\sigma(q)&=& \Delta A_y[-(q+s_y)\log (q+s_y) \nonumber\\
& & + q(1+s_y)\log (1+s_y) +(1-q)s_y\log s_y]\nonumber\\
& & + \Delta A_z[(q+s_z)\log (q+s_z) \nonumber\\
& & - q(1+s_z)\log (1+s_z) -(1-q)s_z\log s_z].\label{eq:keisan1}
\eeqn
Then, a direct computation shows that
\beqn
\sigma^{\prime}(q)
&=& \Delta A_y[-\log (q+s_y) -1 \nonumber\\
& & + (1+s_y)\log (1+s_y) -s_y\log s_y]\nonumber\\
& & + \Delta A_z[\log (q+s_z) +1\nonumber\\
& & - (1+s_z)\log (1+s_z)+s_z\log s_z]\nonumber\\ 
& =& \Delta \left[(A_z-A_y) - \log\frac{(q+s_y)^{A_y}}{(q+s_z)^{A_z}}\right.\nonumber\\
& & \left.+ \log\frac{(1+s_y)^{(1+s_y)A_y}}{(1+s_z)^{(1+s_z)A_z}}
-\log\frac{s_y^{s_yA_y}}{s_z^{s_zA_z}}\right].
\eeqn
Hence, the solution $q =q^*$ of the equation $\sigma^{\prime}(q)=0$ is given by
\beqn\label{eq:keisan3}
 \log\frac{(q^*+s_y)^{A_y}}{(q^*+s_z)^{A_z}} 
 &=& (A_z - A_y) 
+ \log\frac{(1+s_y)^{(1+s_y)A_y}}{(1+s_z)^{(1+s_z)A_z}}\nonumber\\
& & -\log\frac{s_y^{s_yA_y}}{s_z^{s_zA_z}},
\eeqn
which is equivalent to
\beq\label{eq:keisan7}
\frac{(A_yq^*+\lambda_y)^{A_y}}{(A_zq^*+\lambda_z)^{A_z}}
= e^{A_z-A_y}\frac{(A_y+\lambda_y)^{A_y+\lambda_y}}{(A_z+\lambda_z)^{A_z+\lambda_z}}
\frac{\lambda_z^{\lambda_z}}{\lambda_y^{\lambda_y}}.
\eeq
On the other hand, 
\beqn\label{eq:keisanc4}
\sigma(q) &=& \Delta \log\frac{(q+s_z)^{(q+s_z)A_z}}{(q+s_y)^{(q+s_y)A_y}}\nonumber\\
& & + \Delta q\log\frac{(1+s_y)^{(1+s_y)A_y}}{(1+s_z)^{(1+s_z)A_z}}\nonumber\\
& & + \Delta (1-q)\log\frac{s_y^{s_yA_y}}{s_z^{s_zA_z}}\nonumber\\
& =&  \Delta \log\frac{(q+s_z)^{(q+s_z)A_z}}{(q+s_y)^{(q+s_y)A_y}}
+ \Delta \log\frac{s_y^{s_yA_y}}{s_z^{s_zA_z}}\nonumber\\
& & + \Delta q\left(\log\frac{(1+s_y)^{(1+s_y)A_y}}{(1+s_z)^{(1+s_z)A_z}}
-\log\frac{s_y^{s_yA_y}}{s_z^{s_zA_z}}\right)\nonumber\\
&=&  \Delta \log\frac{(q+s_z)^{(q+s_z)A_z}}{(q+s_y)^{(q+s_y)A_y}}
+ \Delta \log\frac{s_y^{s_yA_y}}{s_z^{s_zA_z}}\nonumber\\
& & + \Delta q\left(\log\frac{(q^*+s_y)^{A_y}}{(q^*+s_z)^{A_z}} + A_y - A_z\right),
\eeqn
where we used (\ref{eq:keisan3}) in the last step. Consequently, 
with $q^*_{\Gamma} = \min (q^*, \Gamma)$,
\beqn\label{eq:keisuu1}
\lefteqn{\max_{X:\Exp c(X)\le \Gamma}(I(X;Y)-I(X;Z))}\nonumber\\
&=& \max_{0\le q\le \Gamma}(I(X;Y)-I(X;Z))\nonumber\\
&=&\Delta \log\frac{(q^*_{\Gamma}+s_z)^{(q^*_{\Gamma}+s_z)A_z}}
{(q^*_{\Gamma}+s_y)^{(q^*_{\Gamma}+s_y)A_y}}
+ \Delta \log\frac{s_y^{s_yA_y}}{s_z^{s_zA_z}}\nonumber\\
& & + \Delta q^*_{\Gamma}\left(\log\frac{(q^*+s_y)^{A_y}}{(q^*+s_z)^{A_z}} + A_y - A_z\right).
\eeqn
Since Theorem \ref{teiri:wagcont41} claims that the left-hand side of (\ref{eq:keisuu1})
gives the $\delta$-secrecy capacity per channel use, it is concluded that the 
$\delta$-secrecy capacity 
$\delta$-$C_s(\Gamma)$ per second
is given by  (\ref{eq:karuwaza23}). \QED
%
%

%
%
%
%
\brei\label{rei:wag5}
{\rm
It is easy to check that, in the special case {\em without} cost constraint (i.e., $\Gamma =1$ and hence 
$q^*_{\Gamma}=q^* $),  
 (\ref{eq:karuwaza23}) boils down to
 \beqn\label{eq:karuwaza35}
\delta\mbox{-}C_s(1)
&=& q^*(A_y-A_z) + \log\frac{\lambda_y^{\lambda_y}}{\lambda_z^{\lambda_z}}\nonumber\\
& & +\log\frac{(A_zq^*+\lambda_z)^{\lambda_z}}{(A_yq^*+\lambda_y)^{\lambda_y}},
\eeqn
which coincides with the average criterion formula for the  w-$C_s(1)$ 
as already developed in the {\em continuous time} framework by Laourine and Wagner \cite{wagner} 
with the same equation as (\ref{eq:keisan70}).
As for the definition of  $\mbox{w-}C_s(\Gamma)$, 
see Section \ref{intro-geri1}.
From the security point of view, 
formula (\ref{eq:karuwaza35}) is  stronger than the formula for  $\mbox{w-}C_s(1)$
 as was discussed in Section \ref{intro-geri1}.C, 
though $\delta\mbox{-}C_s(1) = \mbox{w-}C_s(1)$. 
%
\QED
}
\erei
\brei\label{rei:laour1}
{\rm 
Let us quote here the {\em worst case scenario} as demonstrated in \cite{wagner} 
specified by
\[
\frac{\lambda_y}{A_y} = \frac{\lambda_z}{A_z} =s.
\]
In this case, it is shown in \cite{wagner} that $q^*$ is given by
\beq\label{eq:sinario1}
q^* = \frac{(1+s)^{1+s}}{es^s} -s.
\eeq
It is then also easy to verify that (\ref{eq:karuwaza23}) reduces to
\beq\label{eq:delta1}
\delta\mbox{-}C_s(\Gamma) = (A_y-A_z)\left[
\begin{array}{l}
-(q_{\Gamma}+s)\log(q_{\Gamma}+s) +s\log s\\
\qquad\qquad +q_{\Gamma}\left[(1+s)\log(1+s) -s\log s\right]
\end{array}
\right],
\eeq
where 
\[
q_{\Gamma}=\min\left(\frac{(1+s)^{1+s}}{es^s} -s, \Gamma\right).
\]
Moreover, in the particular case with $s=0$ (no dark current), (\ref{eq:delta1}) reduces to 
\beq\label{eq:eiga1}
\delta\mbox{-}C_s(\Gamma)= -(A_y-A_z)q_{\Gamma}\log q_{\Gamma},
\eeq
where
\[
q_{\Gamma}=\min\left(\frac{1}{e}, \Gamma\right).
\]

\QED
}
\erei

\section{Reliability and secrecy functions  of Poisson wiretap channel}\label{intro-poisson-re-se2}
In this section, we consider application of Theorem \ref{teiri:max2} to the Poisson wiretap channel to
evaluate its reliability and secrecy functions.
Here too, as in the previous section, 
we use the same  two-input two-output stationary memoryless channel model specified with the  
 transition probabilities and the  cost constraint with
parameters (\ref{eq;po5}) $\sim$ (\ref{eq:postq1}). In this section we focus on Poisson wiretap channels 
{\em without} concatenation 
(i.e., $V\equiv X$; cf. Remark \ref{chui:vxeq}),
and later in Section \ref{intro-poisson-re-conc1}  extend it  to the case of  Poisson wiretap 
channels {\em with} concatenation.
Also, we assume that the conditions for degradedness 
(\ref{eq:degc1}) and (\ref{eq:degc2}) in Theorem \ref{teiri:degraded1} are satisfied.
%

{\em A. Reliability function}

The first concern in this section is on the behavior of the reliability function for Bob.
Formula (\ref{eq:hana51}) 
of Theorem \ref{teiri:max2} with $q=P_X$ is written as 
\beqn\label{eq:kanna2f}
 \mbox{{\scriptsize m}-}\epsilon_n^B &\le & 6e^{-nF_c(q, R_{B0}, R_{E0},n)}\nonumber\\
 &=& \exp[-n\sup_{r\ge 0}\sup_{0\le \rho \le 1}\left(E_{B0}(\rho, q, r)
- \rho (R_{B0}+R_{E0})+O(1/n)\right)]\nonumber\\
&=& \exp[-n\sup_{r\ge 0}\sup_{0\le \rho \le 1}\left(E_{B0}(\rho, q, r)
- \rho (R_{B0}+R_{E0})\right)+O(1)],
\eeqn
where 
we have set $E_{B0}(\rho, q, r) =\phi(\rho|W_B, q,r)$. 
Let us first evaluate $E_{B0}(\rho, q, r)$.
Taking account of (\ref{eq:finder3}), 
we have 
\beqn\label{eq:exponentb1}
E_{B0}(\rho, q, r)&=&
-\log\left[\sum_{y\in \cY}\left(
  \sum_{x\in \cX}q(x)W_B(y|x)^{\frac{1}{1+\rho}}
e^{r[\Gamma-c(x)]}\right)^{1+\rho}\right]\nonumber\\
&=& -\log\left[\sum_{y=0}^1\left(
  \sum_{x=0}^1q(x)W_B(y|x)^{\frac{1}{1+\rho}}
e^{r[\Gamma-c(x)]}\right)^{1+\rho}\right]\nonumber\\
&=& -\log\sum_{y=0}^1V_y^{1+\rho}-r(1+\rho)\Gamma,
\eeqn
where 
\[
V_y =  \sum_{x=0}^1q(x)W_B(y|x)^{\frac{1}{1+\rho}}
e^{-rx} \quad (y=0,1).
\]
(It should be noted here that in evaluation of (\ref{eq:exponentb1})
Wyner  \cite{wyner1} used  $c(x)-\Gamma$ instead of $\Gamma -c(x)$, which causes 
some subtle irrelevance.) \
With $q=q(1)$, an elementary caluculation 
using (\ref{eq;po5}) and (\ref{eq;po6}) leads, up to the  order $O(\Delta)$,  to 
\beqn\label{eq:expo2}
E_{B0}(\rho, q, r)&=& -r(1+\rho)\Gamma - (1+\rho)\log(1-q +qe^{-r})\nonumber\\
& & +\Delta A_y\left[\frac{(1-q)s_y+qe^{-r}(1+s_y)}{1-q +qe^{-r}}\right]\nonumber\\
& & -\Delta A_y\left[\frac{(1-q)s_y^{\frac{1}{1+\rho}}+qe^{-r}(1+s_y)^{\frac{1}
{1+\rho}}}{1-q +qe^{-r}}\right]^{1+\rho}.
\eeqn
First, in order to  maximize $E_{B0}(\rho, q, r)$ with respect to $r$, set
\[
g(r)
        =  -r(1+\rho)\Gamma - (1+\rho)\log(1-q +qe^{-r}). 
\]
Then,

\[
g^{\prime}%
(r)
= -(1+\rho)\Gamma + (1+\rho)\frac{qe^{-r}}{1-q+qe^{-r}},
\]
\[
g^{\prime\prime}
(r)
= -(1+\rho)\frac{q^2e^{-r}(1-q)}{(1-q +qe^{-r})^2} < 0,
\]
which means that $g  (r)$ is strictly concave. 
It is evident that 
\[
g(0)
=0, \quad g^{\prime}%
(0)=-(1+\rho)(\Gamma -q) \le 0,
\]
where we have used that cost constraint $\Exp c(X)\le \Gamma$ 
is written as  $q\le \Gamma$.
Consequently, we have
\beq\label{eq:expro4}
\max_{r\ge 0}g
(r) =g(0)
= 0,
\eeq
and hence, up to the  order $O(\Delta)$,
\beqn\label{eq:expt1}
\lefteqn{E_{B0}(\rho, q)\stackrel{\Delta}{=}\max_{r\ge 0}E_{B0}(\rho, q, r)=E_{B0}(\rho, q, r=0)}\nonumber\\
& =&  \Delta A_y\left[
(1-q)s_y+(1+s_y)q 
-\left[(1-q)s_y^{\frac{1}{1+\rho}}+q(1+s_y)^{\frac{1}{1+\rho}}\right]
^{1+\rho}\right].\nonumber\\
&= & \Delta A_y\left[ q+s_y -s_y(1+\tau_y q)^{1+\rho}\right],
\eeqn
where
\beq\label{eq:expro7}
\tau_y = \left(1+\frac{1}{s_y}\right)^{\frac{1}{1+\rho}} -1.
\eeq
On the other hand, (\ref{eq:kanna2f}) is rewritten as 
\beq\label{eq:expro8}
 \mbox{{\scriptsize m}-}\epsilon_n^B \le 
  \exp[-n\sup_{0\le \rho \le 1}\left(E_{B0}(\rho, q)
- \rho (R_{B0}+R_{E0})\right)+O(1)].
\eeq
Notice here that  $E_{B0}(\rho, q)
- \rho (R_{B0}+R_{E0})$ in (\ref{eq:expro8}) is the exponent per channel use, so that 
\[
\frac{E_{B0}(\rho, q)
- \rho (R_{B0}+R_{E0})}{\Delta}
\] gives the exponent per second. Therefore,
\[
E_B(\rho,q) = \frac{E_{B0}(\rho, q)}{\Delta},\quad R_B=\frac{R_{B0}}{\Delta}, 
\quad R_E=\frac{R_{E0}}{\Delta}
\]
gives the exponents per second. Thus, taking account of $T=n\Delta$, it turns out that
(\ref{eq:expro8}) 
is equivalent to
\beq\label{eq:expro31}
 \mbox{{\scriptsize m}-}\epsilon_n^B \le 
  \exp[-T\sup_{0\le \rho \le 1}\left(E_B(\rho, q)
- \rho (R_B+R_E)\right)+O(1)],
\eeq
where 
\beq\label{eq:expro32}
 E_B(\rho, q) = A_y\left[ q+s_y -s_y(1+\tau_y q)^{1+\rho}\right].
\eeq
We notice that formula (\ref{eq:expro31}) together with (\ref{eq:expro32}) coincides with that 
established by Wyner \cite{wyner-wire}  for non-wiretap Poisson channels, although 
the ways of derivation are different.

Since $E_B(\rho, q)$ is concave in $\rho$  (cf. Gallager \cite{gall}), the supremum
\[
\sup_{0\le \rho \le 1}\left(E_B(\rho, q)
- \rho (R_B+R_E)\right)
\]
 is specified by the equation:
 \beq\label{eq:expro3M}
 \frac{dE_B(\rho, q)}{d\rho} =R_B+R_E \quad (0\le \rho \le1).
\eeq
Carrying out a direct calculation of the left-hand side of (\ref{eq:expro3M}), it follows that
\beqn\label{eq:para1pS}
\lefteqn{R_B+R_E = \frac{dE_B(\rho, q)}{d\rho}}\nonumber\\
&=& A_ys_y\left[q\left(1+\frac{1}{s_y}\right)^{\frac{1}{1+\rho}}
\frac{(1+\tau_yq)^{\rho}}{1+\rho}
\log \left(1+\frac{1}{s_y}\right) - (1+\tau_yq)^{1+\rho}\log(1+\tau_yq)
\right],\nonumber\\
& & 
\eeqn
which together with (\ref{eq:expro7}) and (\ref{eq:expro32}) gives the parametric representation 
of the reliability function under the maximum criterion $\mbox{{\scriptsize m}-}\epsilon_n^B$ 
with parameter $\rho$.
\bchui\label{chui:expo1}
{\rm 
The function 
\beq\label{eq:poexS}
f_B(R, q) \stackrel{\Delta}{=} \sup_{0\le \rho \le 1}\left(E_B(\rho, q)
- \rho R\right)
\quad (R=R_B+R_E)
\eeq
can be derived by eliminating $\rho$ from (\ref{eq:expro32}) using (\ref{eq:para1pS}), 
and is zero at 
\beqn\label{eq:para1S}
\lefteqn{R_B+R_E =}\nonumber\\
&=& A_ys_y\left[q\left(1+\frac{1}{s_y}\right)
\log \left(1+\frac{1}{s_y}\right) - \left(1+\frac{q}{s_y}\right)\log\left(1+\frac{q}{s_y}\right)
\right]\nonumber\\
& \stackrel{\Delta}{=} & h_B(q)=I(q,W_B)/\Delta,
\eeqn
and $f_B(R_B+R_E, q)$ is  convex and positive in the range: $R_B+R_E < h_B(q)$. \QED
}
\echui


\bigskip

{\em B. Secrecy function}

Let us now turn to the problem of evaluating the secrecy function against Eve.
We proceed in parallel with the above case of reliability function.
Formula (\ref{eq:hana61}) with $q=P_X$
of Theorem \ref{teiri:max2}  is written as 
\beqn\label{eq:kanna2fs}
 \mbox{{\scriptsize m}-}\delta_n^E &\le & 6e^{-nH_c(q, R_{E0},n)}\nonumber\\
 &=& \exp[-n\sup_{0 < \rho < 1}\sup_{r\ge 0}\left(E_{E0}(\rho, q, r)
+\rho R_{E0}+O(1/n)\right)]\nonumber\\
&=& \exp[-n\sup_{0< \rho < 1}\sup_{r\ge 0}\left(E_{E0}(\rho, q, r)
+ \rho R_{E0}\right)+O(1)],
\eeqn
where
we have set $E_{E0}(\rho, q, r) =\phi(-\rho|W_E, q,r)$. 
Let us evaluate $E_{E0}(\rho,q, r).$
Taking account of (\ref{eq;halimeq1}), we have 
\beqn\label{eq:exponentbs1}
E_{E0}(\rho, q, r)&=&
-\log\left[\sum_{z\in \cZ}\left(
  \sum_{x\in \cX}q(x)W_E(z|x)^{\frac{1}{1-\rho}}
e^{r[\Gamma-c(x)]}\right)^{1-\rho}\right]\nonumber\\
&=& -\log\left[\sum_{z=0}^1\left(
  \sum_{x=0}^1q(x)W_E(z|x)^{\frac{1}{1-\rho}}
e^{r[\Gamma-c(x)]}\right)^{1-\rho}\right]\nonumber\\
&=& -\log\sum_{z=0}^1V_z^{1-\rho} -r(1-\rho)\Gamma,
\eeqn
where 
\[
V_z =  \sum_{x=0}^1q(x)W_E(z|x)^{\frac{1}{1-\rho}}
e^{-rx} \quad (z=0,1).
\]
With $q=q(1)$, an elementary caluculation 
using (\ref{eq;po7}) and (\ref{eq;po8}) leads, up to the  order $O(\Delta)$,  to 
\beqn\label{eq:expos2}
E_{E0}(\rho, q, r)&=& -r(1-\rho)\Gamma - (1-\rho)\log(1-q +qe^{-r})\nonumber\\
& & +\Delta A_z\left[\frac{(1-q)s_z+qe^{-r}(1+s_z)}{1-q +qe^{-r}}\right]\nonumber\\
& & -\Delta A_z\left[\frac{(1-q)s_z^{\frac{1}{1-\rho}}+qe^{-r}(1+s_z)^{\frac{1}
{1-\rho}}}{1-q +qe^{-r}}\right]^{1-\rho}.
\eeqn
In order to first maximize $E_{E0}(\rho, q, r)$ with respect to $r$, set
\[
h(r)
        =  -r(1-\rho)\Gamma - (1-\rho)\log(1-q +qe^{-r}). 
\]
Then,
\[
h^{\prime}%
(r)
= -(1-\rho)\Gamma + (1-\rho)\frac{qe^{-r}}{1-q+qe^{-r}},
\]
\[
h^{\prime\prime}
(r)
= -(1-\rho)\frac{q^2e^{-r}(1-q)}{(1-q +qe^{-r})^2} < 0,
\]
which means that $h  (r)$ is strictly concave. 
It is evident that 
\[
h(0)
=0, \quad h^{\prime}%
(0)=-(1-\rho)(\Gamma -q) \le 0.
\]
Consequently, we have
\beq\label{eq:expros4}
\max_{r\ge 0}h(r)
=h(0)
= 0,
\eeq
and hence, up to the order $O(\Delta)$,  
\beqn\label{eq:expts1}
\lefteqn{E_{E0}(\rho, q)\stackrel{\Delta}{=}\max_{r\ge 0}E_{E0}(\rho, q, r)=E_{E0}(\rho, q, r=0)}\nonumber\\
& =&  \Delta A_z\left[
(1-q)s_z+(1+s_z)q 
-\left[(1-q)s_z^{\frac{1}{1-\rho}}+q(1+s_z)^{\frac{1}{1-\rho}}\right]
^{1-\rho}\right].\nonumber\\
&= & \Delta A_z\left[ q+s_z -s_z(1+\tau_z q)^{1-\rho}\right],
\eeqn
where
\beq\label{eq:expros7}
\tau_z = \left(1+\frac{1}{s_z}\right)^{\frac{1}{1-\rho}} -1.
\eeq
On the other hand, (\ref{eq:kanna2fs}) is rewritten as 
\beq\label{eq:expros8}
 \mbox{{\scriptsize m}-}\delta_n^E \le 
  \exp[-n\sup_{0< \rho < 1}\left(E_{E0}(\rho, q)
+ \rho R_{E0}\right)+O(1)].
\eeq
Notice here that  $E_{E0}(\rho, q)
+ \rho R_{E0}$ in (\ref{eq:expros8}) is the exponent per channel use, so that 
\[
\frac{E_{E0}(\rho, q)
+ \rho R_{E0}}{\Delta}
\] 
gives the exponent per second. Therefore,
\[
E_E(\rho,q) = \frac{E_{E0}(\rho, q)}{\Delta},\quad R_E=\frac{R_{E0}}{\Delta} 
\]
gives the exponents per second. Thus, taking account of $T=n\Delta$, it turns out that
(\ref{eq:expros8}) 
is equivalent to
\beq\label{eq:expros31}
 \mbox{{\scriptsize m}-}\delta_n^E \le 
  \exp[-T\sup_{0< \rho < 1}\left(E_E(\rho, q)
+ \rho R_E\right)+O(1)],
\eeq
where 
\beq\label{eq:expros32}
 E_E(\rho, q) = A_z\left[ q+s_z -s_z(1+\tau_z q)^{1-\rho}\right].
\eeq
Since $E_E(\rho, q)$ is concave in $\rho$, the supremum
\[
\sup_{0 < \rho < 1}\left(E_E(\rho, q)
+ \rho R_E\right)
\]
 is specified by the equation;
 \beq\label{eq:expros33}
 -\frac{dE_E(\rho, q)}{d\rho} =R_E \quad (0<\rho <1).
\eeq
Carrying out a direct calculation of the left-hand side of (\ref{eq:expros33}), it follows that
\beqn\label{eq:para1}
\lefteqn{R_E = -\frac{dE_E(\rho, q)}{d\rho}}\nonumber\\
&=& A_zs_z\left[q\left(1+\frac{1}{s_z}\right)^{\frac{1}{1-\rho}}
\frac{(1+\tau_zq)^{-\rho}}{1-\rho}
\log \left(1+\frac{1}{s_z}\right) - (1+\tau_zq)^{1-\rho}\log(1+\tau_zq)
\right],\nonumber\\
& & 
\eeqn
which together with (\ref{eq:expros7}) and (\ref{eq:expros32}) gives the parametric representation 
of the secrecy function 
under the maximum criterion $\mbox{{\scriptsize m}-}\delta_n^E$
with parameter $\rho$.
\bchui\label{chui:expo1T}
{\rm 
The function 
\beq\label{eq:poexST}
f_E(R, q) \stackrel{\Delta}{=} \sup_{0< \rho < 1}\left(E_E(\rho, q)
+ \rho R\right)
\quad (R=R_E)
\eeq
can be derived by eliminating $\rho$ from (\ref{eq:expros32}) using (\ref{eq:para1}), 
and is zero at 
\beqn\label{eq:para1ST}
R_E 
&=& A_zs_z\left[q\left(1+\frac{1}{s_z}\right)
\log \left(1+\frac{1}{s_z}\right) - \left(1+\frac{q}{s_z}\right)\log\left(1+\frac{q}{s_z}\right)
\right]\nonumber\\
& \stackrel{\Delta}{=} & h_E(q)=I(q,W_E)/\Delta,
\eeqn
and $f_E(R_E, q)$ is  convex and positive in the range: $R_E > h_E(q)$.
It should be noted here that the form of the function $f_E(R, q)$ is the same as that of 
$f_B(R, q)$ in (\ref{eq:poexS}) of  Remark \ref{chui:expo1}, while they are positive
 in the opposite directions, i.e., (\ref{eq:para1ST}) and 
 $R_E > h_E(q)$
 correspond 
 to (\ref{eq:para1S}) and 
 $R_B+R_E < h_B(q)$,
respectively.
\QED
}
\echui



\begin{figure}[htbp]
\begin{center}
\includegraphics[width=70mm]{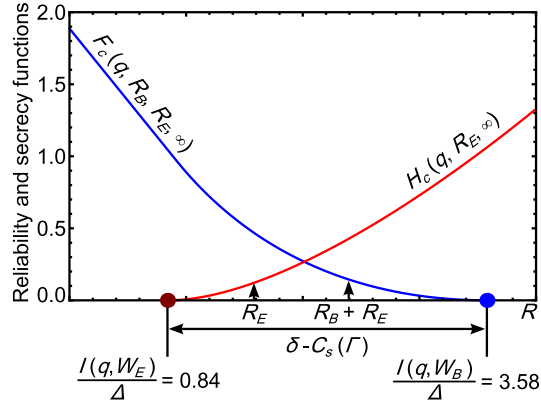}
\end{center}
\caption{Reliability and secrecy functions for Poisson channel ($A_y = 12$, 
$\lambda_y = 0.5$, $A_z = 5$, $\lambda_z = 1.5$, $\Gamma = 0.5$, $q(1)=0.38$).}
\label{fig12}
\end{figure}


\bchui\label{chui:expos3}
{\rm 
As was stated in the previous section, {\em  degradedness} implies {\em more capability}, so that
it holds  in the non-degenerated case that $I(q;W_B) > I(q;W_E)$ for some $q$ owing to the assumed  degradedness, 
which guarantees  that the secrecy function curve
crosses the reliability function curve.
This property enables us to control the tradeoff between reliability and secrecy
(cf. Section \ref{sec:tradeoff}.).
It should be noted here  that in the above arguments the common input probability $q$ is shared 
by both the reliability function and the secrecy function. This implies that maximization 
over $q$ should not be taken 
separately for the reliability function and the secrecy function, but should be taken for 
$I(q;W_B) - I(q;W_E)$ to achieve the $\delta$-secrecy capacity $\delta$-$C_s(\Gamma)$ of the 
wiretap channel, as long as $q$ satisfies the cost constraint $q\le\Gamma$.
A typical case is illustrated in Fig.\ref{fig12}.
\QED
}
\echui
\section{Concatenation for  Poisson wiretap channel}\label{intro-poisson-re-conc1}
In this section, we investigate the effects of  {\em concatenation} 
for performance of Poisson wiretap channels. We first observe 
a basic property ({\em invariance})  of  Poisson wiretap channel under concatenation
(on the basis of Theorem \ref{teiri:max2}  and Theorem \ref{teiri:wagcont41}).
Here too, we use the notation as used in Sections \ref{intro-poisson}, \ref{intro-poisson-re-se2}.
\bteiri\label{teiri:sonsa1}
{\rm
If a Poisson wiretap channel is degraded, i.e.,  (\ref{eq:degc1}) and (\ref{eq:degc2}) 
are satisfied, then its  concatenated Poisson wiretap channel  also 
satisfies the same form of conditions as (\ref{eq:degc1}) and (\ref{eq:degc2}).
\QED
}
\eteiri

{\em Proof}:\ \ 
Set the transition probabilities $P_{X|V}$ of the auxiliary binary channel as 
\beqn
& & a=P_{X|V}(1|1),\quad 1-a=P_{X|V}(0|1);\label{eq:sonsa3}\\
& & b=P_{X|V}(1|0),\quad 1-b=P_{X|V}(0|0),\label{eq:sonsa2}
\eeqn
where we assume that $1\ge a>b \ge 0$.
Then, the transition probabilities of the concatenated channel
$(W^+_B, W^+_E)$ are given by 
 \beqn
W^+_B(1|1) &=& \left[a(A_y+\lambda_y)+(1-a)\lambda_y\right] \Delta
= [aA_y +\lambda_y] \Delta,\label{eq;sonsapo6}\\
W^+_B(1|0) &=&\left[b(A_y+\lambda_y)+(1-b)\lambda_y\right] \Delta
= [bA_y +\lambda_y] \Delta;\label{eq;sonsapo5}
\eeqn
\beqn
W^+_E(1|1) &=& \left[a(A_z+\lambda_z)+(1-a)\lambda_z\right] \Delta
= [aA_z +\lambda_z] \Delta,\label{eq;sonsapo8}\\
W^+_E(1|0) &=&\left[b(A_z+\lambda_z)+(1-b)\lambda_z\right] \Delta
= [bA_z +\lambda_z] \Delta.\label{eq;sonsapo7}\
\eeqn
Notice that the concatenated channel is also a Poisson wiretap channel,
and let the peak power and dark currents of the concatenated channel be denoted by
$A^+_y, A^+_z,$ $\lambda^+_y, \lambda^+_z$, respectively, then we obtain
\beqn
\lambda^+_y &=& W^+_B(1|0)/\Delta= bA_y +\lambda_y,\label{eq;sonsapo10}\\
A^+_y &=& (W^+_B(1|1) - W^+_B(1|0))/\Delta=(a-b)A_y,\label{eq;sonsapo11}
\eeqn
\beqn
\lambda^+_z &=& W^+_E(1|0)/\Delta= bA_z +\lambda_z,\label{eq;sonsapo12}\\
A^+_z &=& (W^+_E(1|1) - W^+_E(1|0))/\Delta=(a-b)A_z,\label{eq;sonsapo13}
\eeqn
which means that concatenation has the effect of  not only attenuating peak powers to a factor 
of $a-b$
but also  augmenting 
 a factor of $b$ to dark currents.
Recall that we have set as
\beq\label{eq:sonsapost1}
s_y = \frac{\lambda_y}{A_y}, \quad s_z = \frac{\lambda_z}{A_z}.
\eeq

According to (\ref{eq:sonsapost1}), set
\beq\label{eq:sonsapost15}
s^+_y = \frac{\lambda^+_y}{A^+_y}, \quad s^+_z = \frac{\lambda^+_z}{A^+_z},
\eeq
then
\beqn
s^+_y &=& \frac{bA_y +\lambda_y}{(a-b)A_y} =\frac{b+s_y}{a-b},
\label{eq:sonsaS1}\\
s^+_z &=& \frac{bA_z +\lambda_y}{(a-b)A_z} =\frac{b+s_z}{a-b}.\label{eq:sonsaS2}
\eeqn
from which it follows that
\beq\label{eq:sonsaS2}
s_y \le s_z \Longleftrightarrow s^+_y \le s^+_z.
\eeq
Moreover, from (\ref{eq;sonsapo11}) and (\ref{eq;sonsapo13}) it follows that
\beq\label{eq:sonsaS3}
A_y \ge A_z \Longleftrightarrow A^+_y \ge A^+_z,
\eeq
which completes the proof.
\QED

Since we are considering the case where the non-concatenated channel $(W_B, W_E)$
satisfies conditions (\ref{eq:degc1}) and (\ref{eq:degc2}), Theorem \ref{teiri:sonsa1} 
ensures that the concatenated channel $(W_B^+, W_E^+)$  also satisfies
these conditions as well.  Therefore, in view of Theorem \ref{teiri:degraded1} 
and Theorem \ref{teiri:degraded2},
 $(W_B^+, W_E^+)$ is more capable, so that we can use the same arguments as were 
 developed in Section \ref{intro-poisson}.
On the other hand,  $p=\Pr\{X=1\}$ is given as 
\beq\label{eq:sonsaT1}
 p=qa+(1-q)b
\eeq
with $q=\Pr\{V=1\}$.
Therefore, solving (\ref{eq:sonsaT1}) with respect to $q$, we see that the problem with
  cost constraint $p\le \Gamma$ ($c(x)=x$)  on $P_X$  is equivalent to
 cost constraint $\Gamma^+$ ($c(v)=v$) on $P_V$  
such that
 \beq\label{eq:Sonsa-302}
 q\le \frac{\Gamma -b}{a-b} \stackrel{\Delta}{=} \Gamma^+,
 \eeq
where $\Gamma \ge b$ is assumed (cf.  Section \ref{intro-geri2}.C with $\overline{c}(1)=a, \overline{c}(0)=b$). 
Thus, based on  (\ref{eq:sonsaS2}) $\sim$ (\ref{eq:Sonsa-302}),
we can develop the same arguments on secrecy capacity as well as reliability/secrecy functions 
as in Sections \ref{intro-poisson} and \ref{intro-poisson-re-se2},
which will be briefly summarized in the sequel.
%

{\em A. Secrecy capacity}

The following theorem is the concatenation counterpart of Theorem \ref{teiri:wyn1}
without concatenation.

\bteiri\label{teiri:sonsa-310}
{\rm
Let $a>b$ and $ \Gamma \ge b$. Then,
the  $\delta$-secrecy capacity with cost constraint
$\delta$-$C^+_s(\Gamma)$ per second of the concatenated Poisson wiretap channel $(W^+_B, W^+_E)$ 
is given by  
\beqn\label{eq:karuwaza23T}
\lefteqn{\delta\mbox{-}C^+_s(\Gamma)}\nonumber\\
&=&
\log\frac{(q^*_{\Gamma}+s^+_z)^{(q^*_{\Gamma}+s^+_z)A^+_z}}
{(q^*_{\Gamma}+s^+_y)^{(q^*_{\Gamma}+s^+_y)A^+_y}}
+  \log\frac{(s^+_y)^{s^+_yA^+_y}}{(s^+_z)^{s^+_zA^+_z}}\nonumber\\
& & +  q^*_{\Gamma}\left(\log\frac{(q^*+s^+_y)^{A^+_y}}{(q^*+s^+_z)^{A^+_z}} + A^+_y - A^+_z\right)
\eeqn
under the maximum criterion (m-$\epsilon_n^B$, m-$\delta_n^E$),
where
$q=q^*$ is the unique solution in $(0,1)$ of the equation:
\beq\label{eq:keisan70T}
\frac{(A^+_yq^*+\lambda^+_y)^{A^+_y}}{(A^+_zq^*+\lambda^+_z)^{A^+_z}}
= e^{A^+_z-A^+_y}\frac{(A^+_y+\lambda^+_y)^{A^+_y+\lambda^+_y}}{(A^+_z
+\lambda^+_z)^{A^+_z+\lambda^+_z}}
\frac{(\lambda^+_z)^{\lambda^+_z}}{(\lambda^+_y)^{\lambda^+_y}},
\eeq
and 
\beq\label{eq:karuwaza9T}
q^*_{\Gamma} = \min (q^*, \Gamma^+).
\eeq
}
\eteiri
{\em Proof}:\ \ 
It is not difficult  to check that $0<q*<1$ as was shown in Section \ref{intro-poisson}.
Then, it suffices to proceed in parallel with the proof of Theorem \ref{teiri:wyn1}.
\QED
\brei\label{chui:wag1T}
{\rm
It is easy to check that, in the special case {\em without} cost constraint 
(i.e., $\Gamma =a$ and hence 
$q^*_{\Gamma}=q^* $),  
 (\ref{eq:karuwaza23T}) reduces to
 \beqn\label{eq:karuwaza35T}
\delta\mbox{-}C^+_s(a)
&=& q^*(A^+_y-A^+_z) + \log\frac{(\lambda^+_y)^{\lambda^+_y}}{(\lambda^+_z)^{\lambda^+_z}}
\nonumber\\
& & +\log\frac{(A^+_zq^*+\lambda^+_z)^{\lambda^+_z}}{(A^+_yq^*
+\lambda^+_y)^{\lambda^+_y}}
\eeqn
with  equation (\ref{eq:keisan70T}). 
\QED
}
\erei



%


%
{\em B. Reliability function}

\bteiri\label{teiri:sonsaK1}
{\rm
The maximum error probability $ \mbox{{\scriptsize m}-}\epsilon_n^B$ for $(W^+_B, W^+_E)$ 
is upper bounded  (with $0\le q\le \Gamma^+$) as 
\beq\label{eq:expro31T}
 \mbox{{\scriptsize m}-}\epsilon_n^B \le 
  \exp[-T\sup_{0\le \rho \le 1}\left(E^+_B(\rho, q)
- \rho (R_B+R_E)\right)+O(1)],
\eeq
where 
\beqn
 E^+_B(\rho, q) &=& A^+_y\left[ q+s^+_y -s^+_y(1+\tau^+_y q)^{1+\rho}\right],\label{eq:expro33}\\
 \tau^+_y &=& \left(1+\frac{1}{s^+_y}\right)^{\frac{1}{1+\rho}} -1.\label{eq:expro34}
\eeqn
Furthermore, the $\rho$ to attain the supremum in (\ref{eq:expro31T}) is specified by 
\beqn\label{eq:para1pSS}
\lefteqn{R_B+R_E = \frac{dE^+_B(\rho, q)}{d\rho}}\nonumber\\
&=& A^+_ys^+_y\left[q\left(1+\frac{1}{s^+_y}\right)^{\frac{1}{1+\rho}}
\frac{(1+\tau^+_yq)^{\rho}}{1+\rho}
\log \left(1+\frac{1}{s^+_y}\right)\right.\nonumber\\
& &  - (1+\tau^+_yq)^{1+\rho}\log(1+\tau^+_yq)
\Biggr],
\eeqn
which together with (\ref{eq:expro33}) and (\ref{eq:expro34}) gives the parametric representation 
of the reliability function with parameter $\rho$.
\QED
}
\eteiri

{\em Proof}:\ \ 
%
%
Since the concatenated channel $(W_B^+, W_E^+)$ also satisfies 
conditions (\ref{eq:degc1}) and (\ref{eq:degc2}) with superscript ``$+$,"
it suffices to replace $A_y, \lambda_y, \tau_y$, $E_B(\rho,q)$ in (\ref{eq:expro7}), 
(\ref{eq:expro32}) and (\ref{eq:para1pS}) by
$A^+_y, \lambda^+_y, \tau^+_y$, $E^+_B(\rho,q)$, respectively.
This proof is actually equivalent to the case with $r=0$ in (\ref{eq:istan1x}) in Theorem \ref{chui:tuika3}.
\QED
%
%

%

%
%

%
{\em C. Secrecy function}

\bteiri\label{teiri:sonsaK1M}
{\rm
The maximum divergence $ \mbox{{\scriptsize m}-}\delta_n^E$ for $(W^+_B, W^+_E)$ 
is upper bounded  (with $0\le q\le \Gamma^+$) as 
\beq\label{eq:expro31TM}
 \mbox{{\scriptsize m}-}\delta_n^E \le 
  \exp[-T\sup_{0< \rho < 1}\left(E^+_E(\rho, q)
+ \rho R_E\right)+O(1)],
\eeq
where 
\beqn
 E^+_E(\rho, q) &=& A^+_z\left[ q+s^+_z -s^+_z(1+\tau^+_z q)^{1-\rho}\right],\label{eq:expro33M}\\
 \tau^+_z &=& \left(1+\frac{1}{s^+_z}\right)^{\frac{1}{1-\rho}} -1.\label{eq:expro34M}
\eeqn
Furthermore, the $\rho$ to attain the supremum in (\ref{eq:expro31TM}) is specified by 
\beqn\label{eq:para1pSSM}
\lefteqn{R_E = -\frac{dE^+_E(\rho, q)}{d\rho}}\nonumber\\
&=& A^+_zs^+_z\left[q\left(1+\frac{1}{s^+_z}\right)^{\frac{1}{1-\rho}}
\frac{(1+\tau^+_zq)^{-\rho}}{1-\rho}
\log \left(1+\frac{1}{s^+_z}\right)\right.\nonumber\\
& &  - (1+\tau^+_zq)^{1-\rho}\log(1+\tau^+_zq)
\Biggr],
\eeqn
which together with (\ref{eq:expro33M}) and (\ref{eq:expro34M}) 
gives the parametric representation 
of the secrecy function with parameter $\rho$.
\QED

}
\eteiri  

{\em Proof}:\ \ 
Since the concatenated channel $(W_B^+, W_E^+)$ also satisfies 
conditions (\ref{eq:degc1}) and (\ref{eq:degc2}) with superscript ``$+$,"
it suffices to replace $A_z, \lambda_z, \tau_z$, $E_E(\rho,q)$ in (\ref{eq:expros7}), 
(\ref{eq:expros32}) and (\ref{eq:para1}) by
$A^+_z, \lambda^+_z, \tau^+_z$, $E^+_E(\rho,q)$, respectively.
This proof is actually equivalent to the case with $r=0$ in (\ref{eq:istan2x}) in Theorem \ref{chui:tuika3}.
\QED

\begin{figure}[htbp]
\begin{center}
\includegraphics[width=70mm]{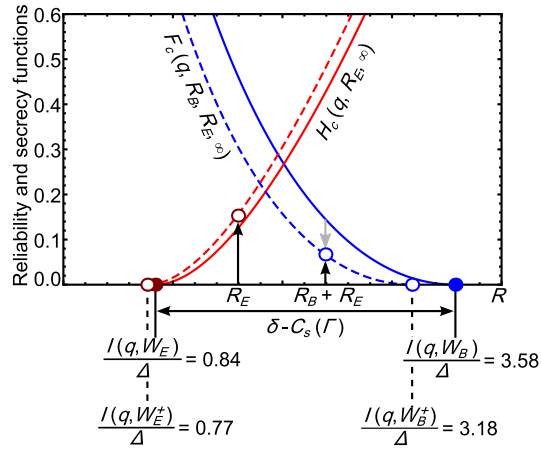}
\end{center}
\caption{Comparison of reliability and secrecy function for non-concatenated (solid line) 
         and concatenated (dash line, $a$=0.98, $b$ = 0.02) Poisson channel ($A_y = 12$, 
         $\lambda_y = 0.5$, $A_z = 5$, $\lambda_z = 1.5$, $\Gamma = 0.5, q(1)=0.38$).}
\label{fig19}
\end{figure}

Typical forms of reliability and secrecy functions of
Poisson wiretap channel with and without concatenation are depicted together in Fig.\ref{fig19}.
%
%
%


%
%
%
%
\section{Secrecy capacity of Gaussian wiretap channel}\label{intro-gaussian-re-se2}
In this section, we first consider application of Theorem \ref{teiri:wagcont41}  to the discrete time 
stationary
memoryless Gaussian wiretap channel to determine the $\delta$-secrecy capacity.
Let the Gaussian wiretap channel be denoted by $(W_B,W_E)$ and the input by $X$, and 
let  $Y,  Z$  be the outputs via channels $W_B,W_E$, respectively,  due to the input $X$, i.e.,
\beqn
Y&=&A_yX+N_y, \label{eq:gauss-out1} \\
Z&=&A_zX+N_z, \label{eq:gauss-out2} 
\eeqn
where $A_y >0, A_z>0$ are positive constants specifying attenuation of signal,
and $N_y, N_z$ are Gaussian {\em additive} noises with 
variances $\sigma_y^2, \sigma_z^2$, respectively. Here, we have 
an analogue of Theorem \ref{teiri:degraded1}:
\bteiri\label{teiri:gdegraded1}
{\rm 
A Gaussian wiretap channel is  (statistically) degraded if
\beq\label{eq:gdegc2}
\frac{\sigma_y}{A_y} \le \frac{\sigma_z}{A_z}.
\eeq
}
\eteiri

{\em Proof}:\ \ 
Set
\[
\tilde{\sigma}^2 = \frac{A_y^2}{A_z^2}\sigma_z^2-\sigma_y^2,
\]
where $\tilde{\sigma} \ge 0$ follows from (\ref{eq:gdegc2}).
Then, there exists   a fictitious Gaussian noise  $\tilde{N}$ with variance $\tilde{\sigma}^2$
 that is 
independent from $N_y$ such that
%
\[
A_yX + N_y +\tilde{N} \simeq A_yX + \frac{A_y}{A_z}N_z = \frac{A_y}{A_z}(A_zX+N_z),
\]
where ``$U \simeq V$" means that $U$ and $V$ are subject to the same statistics.
In view of (\ref{eq:gdegc2}), this means that $A_zX+N_z$ can be obtained by 
adding the fictitious noise $\tilde{N}$ and  attenuating
$A_yX+N_y+\tilde{N}$. \QED

Hereafter, we assume that  condition (\ref{eq:gdegc2}) is satisfied.
Since  degradedness implies more capability (cf. Theorem \ref{teiri:degraded2}),
we can invoke Theorem \ref{teiri:wagcont41} with  cost function $c(x)=x^2$
and cost constraint $\Exp [c(X)] \le \Gamma$ to have
\bteiri\label{teiri:gauss1}
{\rm
The $\delta$-secrecy capacity $\delta$-$C_s(\Gamma)$ of a Gaussian wiretap channel 
under cost constraint $\Gamma$ is given by
\beq\label{eq:gauss-capk}
\delta\mbox{-}C_s(\Gamma) = \frac{1}{2}\log\left(1+\frac{A_y^2\Gamma}{\sigma_y^2}\right)
-\frac{1}{2}\log\left(1+\frac{A_z^2\Gamma}{\sigma_z^2}\right),
\eeq
under the maximum criterion (m-$\epsilon_n^B$, m-$\delta_n^E$).
}
\eteiri
\brei\label{rei:gaussq1}
{\rm
A weak secrecy version of formula (\ref{eq:gauss-capk}) with $A_y=A_z=1$ is found earlier in 
Cheong and Hellman \cite{hellman}: under the average criterion,
\beq\label{eq:gauss-cap10}
\mbox{w-}C_s(\Gamma) = \frac{1}{2}\log\left(1+\frac{\Gamma}{\sigma_y^2}\right)
-\frac{1}{2}\log\left(1+\frac{\Gamma}{\sigma_z^2}\right).
\eeq
}
\erei

{\em Proof of Theorem \ref{teiri:gauss1}}: 

Define  the differential entropy for probability density function $f(u)$ by
\[
h(f) = -\int f(u)\log f(u) du. 
\]
Then,
\beqn\label{eq:gaussianeq1}
\lefteqn{I(X;Y) - I(X;Z)}\nonumber\\
 &=& h(Y) -h(Z) -h(Y|X) +h(Z|X)\nonumber\\
&=&  h(Y) -h(Z) -h(N_y)+h(N_z)\nonumber\\
&=& h(Y) -h(Z) -\frac{1}{2}\log\sigma_y^2 +\frac{1}{2}\log\sigma_z^2\nonumber\\
&=& h\left(\frac{Y}{A_y}\right) - h\left(\frac{Z}{A_z}\right) +\log(A_y/A_z) 
-\frac{1}{2}\log\sigma_y^2 +\frac{1}{2}\log\sigma_z^2.
\eeqn
%
%
We now observe the following equivalence: 
\beq\label{eq:gauss-cap12}
\max_{X:\Exp c(X)\le\Gamma}(I(X;Y) - I(X:Z)) \quad \Longleftrightarrow
\max_{X:\Exp c(X)\le\Gamma}\left( h\left(\frac{Y}{A_y}\right) - h\left(\frac{Z}{A_z}\right)\right).
\eeq
On the other hand, Liu and Viswanath \cite{vis} guarantees  that the maximization 
on the right-hand side is
attained by a Gaussian density $P_X$ with variance $\sigma^2\le \Gamma$.
It is then easy to check that 
\beqn\label{eq:gascom2}
g({\Gamma}) &\stackrel{\Delta}{=} & 
\max_{X:\Exp c(X)\le\Gamma}\left( h\left(\frac{Y}{A_y}\right) - h\left(\frac{Z}{A_z}\right)\right)
\nonumber\\
&=&\frac{1}{2} \max_{\sigma^2\le \Gamma}\left[\log\left(\sigma^2 
+\frac{\sigma_y^2}{A_y^2}\right)
-\log\left(\sigma^2 +\frac{\sigma_z^2}{A_z^2}\right)\right]\nonumber\\
&=& \frac{1}{2}\log\left(\Gamma +\frac{\sigma_y^2}{A_y^2}\right)
-\frac{1}{2}\log\left(\Gamma +\frac{\sigma_z^2}{A_z^2}\right),
\eeqn
where in the last step we have used (\ref{eq:gdegc2}). Susbtituting (\ref{eq:gascom2})
into (\ref{eq:gaussianeq1}) and rearranging it, we eventually obtain
\beq\label{eq:gauss-secrecy1}
\max_{X:\Exp c(X)\le\Gamma}(I(X;Y) - I(X:Z)) =
\frac{1}{2}\log\left(1+\frac{A_y^2\Gamma}{\sigma_y^2}\right)
-\frac{1}{2}\log\left(1+\frac{A_z^2\Gamma}{\sigma_z^2}\right),
\eeq
which together with Theorem \ref{teiri:wagcont41} concludes  
Theorem \ref{teiri:gauss1}.
\QED
\section{Reliability and secrecy functions of Gaussian wiretap channel}\label{intro-gaussian-re-se4}
In this section, we consider application of Theorem \ref{teiri:max2} 
to the Gaussian wiretap channel to
evaluate its reliability and secrecy functions. To this end, it is convenient here  to use,
according to  (\ref{eq:hana51}) and (\ref{eq:hana61}), formulas
\beqn
\mbox{{\scriptsize m}-}\epsilon_n^B &\le &
\exp[-n\sup_{s\ge 0}\sup_{0\le  \rho \le 1}\left(E_B(\rho, q, s)
- \rho (R_B+R_E))+O(1)\right],
 \label{eq:gall-expo-1}\\
\mbox{{\scriptsize m}-}\delta_n^E &\le &
\exp[-n\sup_{s\ge 0}\sup_{0< \rho < 1}\left(E_E(\rho, q, s)
+\rho R_E)+O(1\right)],
 \label{eq:gall-expo-2}
\eeqn
where 
\beqn
E_B(\rho, q, s)&=&
-\log\left[\int_{y}\left(
  \int_{x}q(x)W_B(y|x)^{\frac{1}{1+\rho}}
e^{s[\Gamma-c(x)]}dx\right)^{1+\rho}dy\right],\label{eq:gall-expo-3}\\
E_E(\rho, q, s)&=&
-\log\left[\int_{z}\left(
  \int_{x}q(x)W_E(z|x)^{\frac{1}{1-\rho}}
e^{s[\Gamma-c(x)]}dx\right)^{1-\rho}dz\right].\label{eq:gall-expo-4}
\eeqn
\bchui\label{chui:gala-sec1}
{\rm
These formulas (\ref{eq:gall-expo-1}) $\sim$  (\ref{eq:gall-expo-4}) 
are the continuous alphabet non-concatenated versions of  
(\ref{eq:hana51}) and 
(\ref{eq:hana61}) in Theorem \ref{teiri:max2}
 (cf. Remark \ref{chui:vxeq}). 
}
\QED
\echui
%
%
%

{\em A. Reliability function}

%
%
We  first insert the transition probability density of the Gaussian channel $W_B$:
\beq\label{eq:ahah1}
W_B(y|x) = \frac{1}{\sqrt{2\pi\sigma^2_B}}
\exp\left[-\frac{(y-A_yx)^2}{2\sigma^2_B}\right]
\eeq
and the input distributen for $X$:
\beq\label{eq:ahah2}
q(x)  = \frac{1}{\sqrt{2\pi \Gamma}}
\exp\left[-\frac{x^2}{2\Gamma}\right]
\eeq
into (\ref{eq:gall-expo-3}) to have 
\beqn\label{eq:ahah3}
\lefteqn{E_B(\rho, q, s)}\nonumber\\
&=& -s(1+\rho)A_y^2\Gamma +\frac{1}{2}\log(1+2sA_y^2\Gamma) +\frac{\rho}{2}
\log\left[1+2sA_y^2\Gamma +\frac{A_y^2\Gamma}{(1+\rho)\sigma^2_B}\right],\nonumber\\
\eeqn
where $s\ge 0$ is an arbitrary constant. 
Set
\beqn
A_B &=& \frac{A_y^2\Gamma}{\sigma_B^2},\label{hahaq1}\\
\beta_B &=& 1+2sA_y^2\Gamma + \frac{A_B}{(1+\rho)},\label{hahaq2}
\eeqn
where $\beta_B$ ranges as
\beq\label{eq:station1}
 1+\frac{A_B}{1+\rho} \le \beta_B < +\infty.
\eeq
Use (\ref{hahaq1}) and (\ref{hahaq2}) to eliminate $\Gamma, \sigma_B^2$ and $s$
from (\ref{eq:ahah3}), and consider $E_B(\rho, q, s)$  as a function 
$E_B(A_B, \beta_B, \rho)$ of $A_B, \beta_B, \rho$, 
then 
\beqn\label{eq:exgausre1}
\lefteqn{E_B(A_B, \beta_B, \rho)}\nonumber\\
&=& \frac{1}{2}\left[(1-\beta_B)(1+\rho) +A_B +
\log\left(\beta_B - \frac{A_B}{1+\rho}\right)  + \rho\log \beta_B\right].
\eeqn
Hence,
\beq\label{eq:exgausre4}
\frac{dE_B}{d\beta_B}=\frac{1}{2}\left[-(1+\rho) + \frac{1+\rho}{\beta_B(1+\rho) - A_B}
+\frac{\rho}{\beta_B}\right].
\eeq
Notice that the right-hand side of (\ref{eq:exgausre4}) is decreasing in $\beta_B$ because
$\beta_B(1+\rho) - A_B> 1+\rho$ owing to (\ref{eq:station1}) and  that 
\[
\frac{dE_B}{d\beta_B}<0 \mbox{\ at\ }  \beta_B = 1+ \frac{A_B}{1+\rho}\quad \mbox{and} \ 
\frac{dE_B}{d\beta_B}<0 \mbox{\ when\ } \beta_B \to +\infty.
\]
Therefore, $E_B$ has the maximum value at $\beta_B = 1+ \frac{A_B}{1+\rho}$, i.e.,
\beq\label{eq:exgausre5}
E_B(\rho) \stackrel{\Delta}{=} \max_{\beta_B \ge 1+ \frac{A_B}{1+\rho}}E_B(A_B, \beta_B, \rho)
=\frac{\rho}{2}\log\left(1+ \frac{A_B}{1+\rho}\right)\quad (0\le \rho \le 1).
\eeq
On the other hand, $E_B(\rho) -\rho (R_B+R_E)$ has a stationary point with respect to $\rho$, i.e.,
\beqn\label{eq:exgausre10}
\lefteqn{\frac{\partial (E_B(\rho)-\rho(R_B+R_E))}{\partial \rho}}\nonumber\\
&=& \frac{1}{2}\log\left(1+ \frac{A_B}{1+\rho}\right)
-\frac{\rho A_B}{2(1+\rho)(1+\rho +A_B)} -(R_B+R_E) =0.\nonumber\\
& & 
\eeqn
Hence, 
\beq\label{eq:exgausre13}
R_B+R_E = \frac{1}{2}\log\left(1+ \frac{A_B}{1+\rho}\right)
-\frac{\rho A_B}{2(1+\rho)(1+\rho +A_B)}.
\eeq
As a consequence, by means of (\ref{eq:exgausre5}) and (\ref{eq:exgausre13}), we obtain
 \beqn\label{eq:exgausre12}
E_B(R_B, R_E) &\stackrel{\Delta}{=} & E_B(\rho)-\rho(R_B+R_E)\nonumber\\
&=& \frac{\rho^2 A_B}{2(1+\rho)(1+\rho +A_B)}.
\eeqn
Thus, we have
\bteiri[Reliability function]\label{teiri;chol-S}
{\rm
The reliability function $E_B(R_B, R_E)$ of a Gaussian wiretap channel 
under the maximum criterion $\mbox{{\scriptsize m}-}\epsilon_n^B$
is given 
by the following parametric representation with $0\le \rho \le 1$:
\beqn
E_B(R_B, R_E) & = & \frac{\rho^2 A_B}{2(1+\rho)(1+\rho +A_B)},\label{eq:kom1}\\
R_B+R_E &=& \frac{1}{2}\log\left(1+ \frac{A_B}{1+\rho}\right)
-\frac{\rho A_B}{2(1+\rho)(1+\rho +A_B)}
\eeqn
for $R_{H,c} \le R_B+R_E \le \frac{1}{2}\log\left(1+A_B\right)$ with
\beq\label{eq:moudamewag1}
R_{H,c} \equiv \frac{1}{2}\log\left(1+\frac{A_B}{2}\right)-\frac{A_B}{4(2+A_B)},
\eeq
whereas, for $0\le R_B+R_E \le R_{H,c}$,
\beq\label{eq:loko1}
E_B(R_B, R_E)=\frac{1}{2}\log\left(1+\frac{A_B}{2}\right) -(R_B+R_E).
\eeq
}
\QED
\eteiri

So far we have established the formula for reliability function based 
on upper bound (\ref{eq:smooth1}).
In contrast with Theorem \ref{teiri;chol-S}, 
Gallager \cite{gall} has derived another reliability function 
based on upper bound (\ref{eq:smooth1m}), leading to the exponent formula 
\beq\label{eq:kondo1}
E_B(\rho, q, s)=
-\log\left[\int_{y}\left(
  \int_{x}q(x)W_B(y|x)^{\frac{1}{1+\rho}}
e^{s[c(x-\Gamma)]}dx\right)^{1+\rho}dy\right]
\eeq
instead of (\ref{eq:gall-expo-3}). It should be noted here that in (\ref{eq:kondo1})
$c(x) -\Gamma$ appears instead of $\Gamma - c(x)$. Then, we have 
\bteiri[Reliability function: Gallager]\label{teiri:gauss-expo1}
{\rm The reliability function of a Gaussian wiretap channel 
under the maximum criterion $\mbox{{\scriptsize m}-}\epsilon_n^B$
is given by 
\beqn\label{eq:exgausre22}
\lefteqn{E_B(R_B, R_E)}\nonumber\\
&=& \frac{A_B}{4\beta_B}\left[(\beta_B +1)-(\beta_B-1)\sqrt{1+\frac{4\beta_B}{A_B(\beta_B-1)}}\right]
\nonumber\\
& & +\frac{1}{2}\log\left[\beta_B- \frac{A_B(\beta_B-1)}{2}
\left(\sqrt{1+\frac{4\beta_B}{A_B(\beta_B-1)}}-1\right)\right]
\eeqn
with $\beta_B=e^{2(R_B+R_E)}$.
Formula (\ref{eq:exgausre22}) is valid in the range of $R=R_B+R_E$ as follows:
\beq\label{eq:exgausre25}
R_{G,c} \equiv \frac{1}{2}\log\left[\frac{1}{2} + \frac{A_B}{4} +\frac{1}{2}
\sqrt{1+\frac{A_B^2}{4}}\right]\le R\le \frac{1}{2}\log(1+A_B).
\eeq
For $R$ less than the left-hand side of (\ref{eq:exgausre25}), we must choose $\rho=1$
yielding
\beq\label{eq:exgausre26}
E_B(R_B, R_E) = 1-\beta_B +\frac{A_B}{2}+\frac{1}{2}
\log\left(\beta_B-\frac{A_B}{2}\right) +\frac{1}{2}\log\beta_B -(R_B+R_E),
\eeq
where 
\[
\beta_B = \frac{1}{2}\left[1+\frac{A_B}{2} + \sqrt{1+\frac{A_B^2}{4}}\right].
\]
\QED
}
\eteiri

\begin{figure}[htbp]
\begin{center}
\includegraphics[width=70mm]{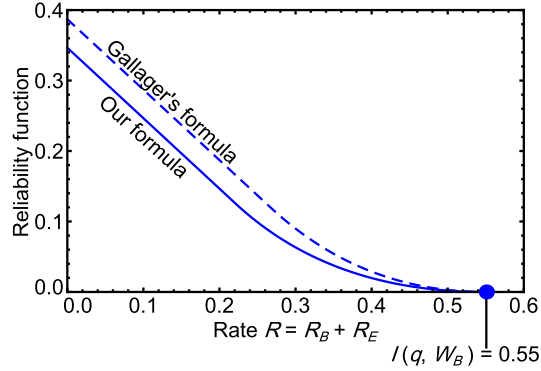}
\end{center}
\caption{Comparison of reliability function for Gaussian channel derived in thist paper (solid line)
and derived by Gallager (dashed line) ($A_y = 1$, $\sigma_y = 0.5$, $\Gamma = 0.5$).}
\label{fig13}
\end{figure}

Two reliability functions derived in the above are depicted in Fig.\ref{fig13}. 
Also, two critical rates $R_{H,c}, R_{G,c}$ defined  in (\ref{eq:moudamewag1}) 
and (\ref{eq:exgausre25}) have the following relation:
\bhodai[Critical rates]\label{hodai:critical}
\beq\label{eq:crit}
R_{H,c} \le R_{G,c}\quad \mbox{for all $A_B$}.
\eeq
\ehodai

{\em Proof:}\  \  Set $g(A_B) = R_{G,c}  - R_{H,c}$. Then,
\[
g^{\prime}(A_B) = \frac{(4+2A_B)\sqrt{1+\frac{A_B^2}{4}}+(4A_B+7A_B^2 +A_B^3-4)}
{2(2+A_B)^2\left((4+2A_B)\sqrt{1+\frac{A_B^2}{4}}+(4+A_B^2)\right)}\ge 0
\]
holds for all $A_B$, which together with $g(0)=0$ yields (\ref{eq:crit}).
\QED

%

%
%
{\em B. Secrecy function}

In this subsection, we evaluate the right-hand side of (\ref{eq:gall-expo-2}) on the secrecy function.
The arguments here proceed in parallel with those in the previous subsection with due modifications
and $-\rho$ instead of $\rho$.
Here too, we insert the transition probability density of the Gaussian channel $W_E$:
\beq\label{eq:ahah11s}
W_E(z|x) = \frac{1}{\sqrt{2\pi\sigma^2_E}}
\exp\left[-\frac{(z-A_zx)^2}{2\sigma^2_E}\right]
\eeq
and the input disturbution for $X$:
\beq\label{eq:ahah12s}
q(x)  = \frac{1}{\sqrt{2\pi \Gamma}}
\exp\left[-\frac{x^2}{2\Gamma}\right]
\eeq
into (\ref{eq:gall-expo-4}) to have 
\beqn\label{eq:ahah3s}
\lefteqn{E_E(\rho, q, s)}\nonumber\\
&=& -s(1-\rho)A_z^2\Gamma +\frac{1}{2}\log(1+2sA_z^2\Gamma) -\frac{\rho}{2}
\log\left[1+2sA_z^2\Gamma +\frac{A_z^2\Gamma}{(1-\rho)\sigma^2_E}\right],\nonumber\\
\eeqn
where $s\ge0$ is an arbitrary constant.
Here we set
\beqn
A_E &=& \frac{A_z^2\Gamma}{\sigma_E^2},\label{hahaq1s}\\
\beta_E &=& 1+2sA_z^2\Gamma + \frac{A_E}{1-\rho},\label{hahaq2s}
\eeqn
then
 $\beta_E$ ranges as
\beq\label{eq:station1s}
1+\frac{A_E}{1-\rho}\le \beta_E <+\infty.
\eeq
%
%
%
Use (\ref{hahaq1s}) and (\ref{hahaq2s}) to eliminate $\Gamma, \sigma_E^2$ and $s$
from (\ref{eq:ahah3s}), and consider $E_E(\rho, q, s)$  as a function 
$E_B(A_E, \beta_E, \rho)$ of $A_B, \beta_E, \rho$, 
then 
\beqn\label{eq:exgausre1s}
\lefteqn{E_E(A_E, \beta_E, \rho)}\nonumber\\
&=& \frac{1}{2}\left[(1-\beta_E)(1-\rho) +A_E +
\log\left(\beta_E - \frac{A_E}{1-\rho}\right)  - \rho\log \beta_E\right].
\eeqn
A stationary point with respect to $\beta_E$ (and hence also with respect to $s$)  is specified by
\beq\label{eq:SSexgausre4}
\frac{dE_E}{d\beta_E}=\frac{1}{2}\left[-(1-\rho) + \frac{1-\rho}{\beta_E(1-\rho) - A_E}
-\frac{\rho}{\beta_E}\right]=0.
\eeq
Notice that the right-hand side of (\ref{eq:SSexgausre4}) is decreasing in $\beta_E$ because
$\beta_E(1-\rho) - A_E\ge 1-\rho$ owing to (\ref{eq:station1s}) and  that 
\[
\frac{dE_E}{d\beta_E}>0 \mbox{\ at\ } \beta_E =1+ \frac{A_E}{1-\rho}\quad \mbox{and} \quad 
\frac{dE_E}{d\beta_E}<0 \mbox{\ when\ } \beta_E \to +\infty.
\]
Therefore, equation (\ref{eq:SSexgausre4}) has the unique solution for $\beta_E$, i.e.,
\beq\label{eq:SSexgausre5}
\beta_E = \frac{1}{2}\left(1+\frac{A_E}{1-\rho}\right)\left[1 +\sqrt{1 +\frac{4A_E\rho}{(1-\rho +A_E)^2}}
\right].
\eeq
On the other hand, $E_E+\rho R_E$ has a stationary point with respect to $\rho$, i.e.,
\beqn\label{eq:SSexgausre10}
\lefteqn{\frac{\partial (E_E+\rho R_E)}{\partial \rho}}\nonumber\\
&=&- \frac{1}{2}\left[1-\beta_E + \frac{\beta_E}{\beta_E (1-\rho) -A_E}
-\frac{1}{1-\rho}+ \log \beta_E\right] +R_E =0.\nonumber\\
& & 
\eeqn
From (\ref{eq:SSexgausre4}) and (\ref{eq:SSexgausre10}), it follows that
\beq\label{eq:SSexgausre13}
R_E = \frac{1}{2}\log\beta_E.
\eeq
Furthermore, combining (\ref{eq:exgausre1s}) with (\ref{eq:SSexgausre13}), we obtain
 \beqn\label{eq:SSexgausre12}
\lefteqn{E_E(R_E)\stackrel{\Delta}{=} E_E+\rho R_E}\nonumber\\
&=& \frac{1}{2}\left[(1-\beta_E)(1-\rho) +A_E +
\log\left(\beta_E - \frac{A_E}{1-\rho}\right) \right].
\eeqn

On the other hand, equation (\ref{eq:SSexgausre4}) can be solved for $\rho$ as follows:
\beq\label{eq:SSexgausre5z}
1-\rho = \frac{A_E}{2\beta_E}\left[1+\sqrt{1+\frac{4\beta_E}{A_E(\beta_E-1)}}\right],
\eeq
which inserted into (\ref{eq:SSexgausre12}) yields the following theorem:
\bteiri[Secrecy function]\label{teiri:SSgauss-expo1}
{\rm The secrecy function of a Gaussian wiretap channel 
under the maximum criterion $\mbox{{\scriptsize m}-}\delta_n^E$
is given by 
\beqn\label{eq:SSexgausre22}
\lefteqn{E_E(R_E)}\nonumber\\
&=& \frac{A_E}{4\beta_E}\left[(\beta_E +1)-(\beta_E-1)\sqrt{1+\frac{4\beta_E}{A_E(\beta_E-1)}}\right]
\nonumber\\
& & +\frac{1}{2}\log\left[\beta_E- \frac{A_E(\beta_E-1)}{2}
\left(\sqrt{1+\frac{4\beta_E}{A_E(\beta_E-1)}}-1\right)\right]
\eeqn
with $\beta_E=e^{2R_E}$.
\QED
}
\eteiri

\bchui\label{chui:gall-exp2}
{\rm
It should be noted that the form of the function in (\ref{eq:exgausre22}) is 
the same as that  in (\ref{eq:SSexgausre22}), but the ranges where they are valid 
are opposite, i.e., 
formula (\ref{eq:SSexgausre22}) is valid in the range of $R_E$:
\beq\label{eq:SSexgausre25}
R_E \ge \frac{1}{2}\log (1+A_E),
\eeq
where parameter $\rho=0$ corresponds to $R_E= \frac{1}{2}\log (1+A_E)$
and $\rho\to 1$ corresponds to $R_E\to +\infty$, whereas 
(\ref{eq:exgausre22}) along with (\ref{eq:exgausre26}) is valid 
when $R_B+R_E \le  \frac{1}{2}\log (1+A_B)$.
\QED
}
\echui

\begin{figure}[htbp]
\begin{center}
\includegraphics[width=70mm]{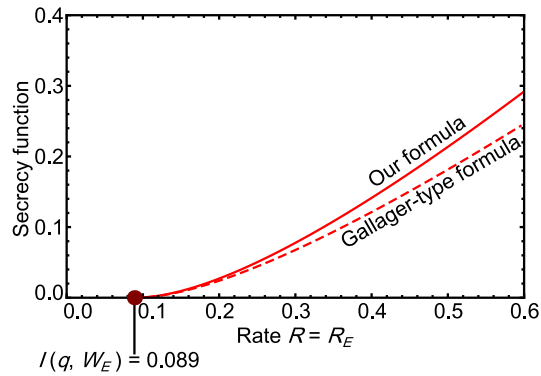}
\end{center}
\caption{Comparison of secrecy function for Gaussian channel with formula derived in this paper (solid line) and Gallager-type formula (dashed line) ($A_z = 0.5$, $\sigma_z = 0.8$, $\Gamma = 0.5$).}
\label{fig14}
\end{figure}

Now in view of Theorem \ref{teiri:gauss-expo1}, one may  be tempted to derive the secrecy function
based on upper bound (\ref{eq:smooth1m}), leading to the exponent formula
\beq\label{kongo2}
E_E(\rho, q, s)=
-\log\left[\int_{z}\left(
  \int_{x}q(x)W_E(z|x)^{\frac{1}{1-\rho}}
e^{s[c(x)-\Gamma]}dx\right)^{1-\rho}dz\right]\label{eq:gall-expo-sec}
\eeq
instead of (\ref{eq:gall-expo-4}). It should be noted here that in (\ref{eq:gall-expo-sec})
$c(x) -\Gamma$ appears instead of $\Gamma - c(x)$.
Let us see what happens in this case.
It is first straightforward to check that (\ref{kongo2}) is developed as 
\beqn\label{eq:ahah3sH}
\lefteqn{E_E(\rho, q, s)}\nonumber\\
&=& s(1-\rho)A_z^2\Gamma +\frac{1}{2}\log(1-2sA_z^2\Gamma) -\frac{\rho}{2}
\log\left[1-2sA_z^2\Gamma +\frac{A_z^2\Gamma}{(1-\rho)\sigma^2_E}\right]\nonumber\\
\eeqn
with
\beqn
A_E &=& \frac{A_z^2\Gamma}{\sigma_E^2},\\
\beta_E &=& 1-2sA_z^2\Gamma + \frac{A_E}{1-\rho},
\eeqn
where it is evident that $\beta_E$ ranges as 
\[
\frac{A_E}{1-\rho} < \beta_E \le 1+\frac{A_E}{1-\rho}.
\]
As was shown in the proof of Theorem \ref{teiri;chol-S}, (\ref{kongo2}) is rewritten as a function of 
$A_E, \beta_E, \rho$ as follows:
\beqn\label{eq:kondou3}
\lefteqn{E_E(A_E, \beta_E, \rho)}\nonumber\\
&=& \frac{1}{2}\left[(1-\beta_E)(1-\rho) +A_E +
\log\left(\beta_E - \frac{A_E}{1-\rho}\right)  - \rho\log \beta_E\right].
\eeqn
Then, it is not difficult to verify that
\beqn\label{eq:kondou5}
E_E(\rho)&\stackrel{\Delta}{=}&\max_{\frac{A_E}{1-\rho} < \beta_E \le 1+\frac{A_E}{1-\rho}}
E_E(A_E, \beta_E, \rho)\nonumber\\
&=&E_E(A_E, 1+\frac{A_E}{1-\rho}, \rho)\nonumber\\
&=&-\frac{\rho}{2}
\log \left(1+\frac{A_E}{1-\rho}\right).
\eeqn
Moreover, the equation 
\[
\frac{\partial (E_E(\rho)+\rho R_E)}{\partial \rho}=0
\]
yields
\beq\label{eq:hontoni1}
R_E= \frac{1}{2}
\log \left(1+\frac{A_E}{1-\rho}\right)
+\frac{\rho A_E}{2(1-\rho)(1-\rho +A_E)}.
\eeq
Then, from (\ref{eq:kondou5}) and (\ref{eq:hontoni1}) it follows that
\beq
E_E(R_E) \stackrel{\Delta}{=} E_E(\rho) +\rho R_E
= \frac{\rho^2A_E}{2(1-\rho)(1-\rho + A_E)}.
\eeq
Thus, we have
\bteiri[Secrecy function: Gallager-type]\label{teiri;chol-T}
{\rm
The secrecy function $E_B(R_B, R_E)$ of a Gaussian wiretap channel 
under the maximum criterion $\mbox{{\scriptsize m}-}\delta_n^E$
is given 
by the following parametric representation with $0\le \rho \le 1$:
\beqn
E_E(R_E) & = & \frac{\rho^2 A_E}{2(1-\rho)(1-\rho +A_E)},\label{eq:kom1G}\\
R_E &=& \frac{1}{2}\log\left(1+ \frac{A_E}{1-\rho}\right)
+\frac{\rho A_E}{2(1-\rho)(1-\rho +A_E)}
\eeqn
for $R_E \ge \frac{1}{2}\log (1+A_E)$.
}
\QED
\eteiri
Two secrecy functions derived in the above are depicted in Fig.\ref{fig14}. 

\begin{figure}[htbp]
\begin{center}
\includegraphics[width=70mm]{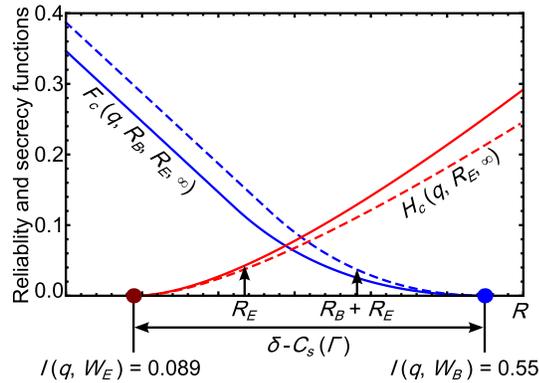}
\end{center}
\caption{Reliability and secrecy functions: 
comparison of Gallager-type formula (dashed line) and our forumla (solid line) 
for Gaussian channel ($A_y = 1$,  $\sigma_y = 0.5$, $A_z = 0.5$, $\sigma_z = 0.8$, $\Gamma = 0.5.
$).}
\label{fig15}
\end{figure}

\begin{figure}[htbp]
\begin{center}
\includegraphics[width=60mm]{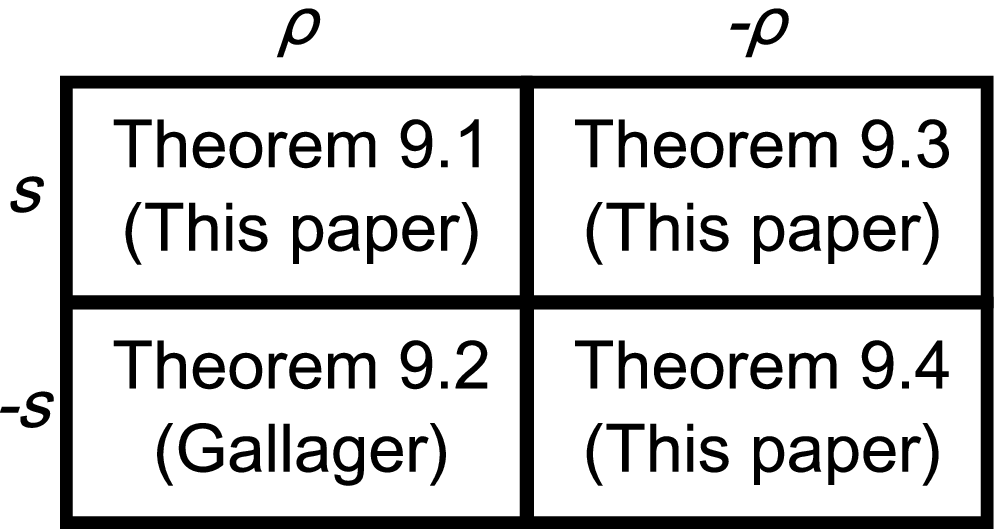}
\end{center}
\caption{The reliability function in Theorem 9.1 (this paper) has the same form as that of the secrecy function in Theorem 9.4 (this paper: Gallager-type),
         whereas the secrecy function in Thoerm 9.3 (this paper) has the same form as that of the reliability function in Theorem 9.2 (Gallager).}
\label{fig16}
\end{figure}

\bchui
{\rm
 In Fig.\ref{fig15} we see that as for the reliability function 
Gallager bound outperforms our bound, whereas as for the secrecy function our bound 
outperforms Gallager-type bound. It is interesting to observe a kind of dualities holding among 
Theorem \ref{teiri;chol-S} $\sim$ Theorem \ref{teiri;chol-T},
which is illustrated in Fig. \ref{fig16}.
\QED
}
\echui

%
%
%
%
%
%
\section{Concluding remarks}\label{conc-remark}
So far we have established  the $\delta$-secrecy capacity with cost constraint
(in the strongest and maximized secrecy sense) 
as well as the pair of reliability and secrecy functions
for the general wiretap channel, and also for 
the stationary memoryless wiretap channel such as 
binary symmetric wiretap channels (BSC), 
Gaussian wiretap channels and Poisson wiretap-channels.
The key concept of the pair of reliability exponent function and secrecy exponent function
has played the crucial role  throughout in this paper.
\\
\ \
Subsequently, we have introduced the formula for the $\delta$-secrecy capacity as the strongest one
among others, which was invoked in many places
 in this paper when cost constraint is considered.
Incidentally, superiority of the maximum secrecy criterion to the average secrecy criterion was demonstrated.
\\
\ \
Next, we have investigated in details one of  typically important channels: the Poisson wiretap channel, whose 
  secrecy-theoretic features have been clarified again from the viewpoint of the pair of reliability and secrecy functions,
  where the formula for the $\delta$-secrecy capacity also naturally followed from the same point of view.
 \\
\ \
  Similarly, also for the Gaussian wiretap channel it was possible to establish the $\delta$-secrecy capacity
  and  the pair of reliability and secrecy functions as well, where we had
   four formulas for reliability and secrecy functions depending 
on different upper bounding techniques
on the characteristic function $\chi (\ssx)$ to ensure to satisfy the cost constraint: one of them is
due to Gallager \cite{gall} and  the other three are demonstrated for the first time in this paper.
These  were shown to have  two-folded  dualities (cf. Fig. \ref{fig16}).
An open problem is left here to make clear the reason.
The $\delta$-secrecy capacity formula for the Gaussian wiretap channel 
under the maximum criterion  was shown to be  stronger than that of Cheong and Hellman \cite{hellman}
from the viewpoint of secrecy.
\\
\ \
Moreover, we have introduced the concept of {\em concatenation}
 in order to expand performance of the wiretap channel. 
Four ways to control the tradeoff 
between reliability and secrecy were shown to be possible  on the basis of 
{\em rate shifting},  {\em rate exchange},
{\em concatenation},
 and {\em change of cost constraint}, respectively.
\\
\ \
Interestingly enough, it turned out that cost constraint $\Gamma$ (with cost $c(x)$) on  $P_{X^n}$ 
  of the concatenated channel
  $(W^{n+}_B, W^{n+}_E)$ is equivalent to cost constraint $\Gamma$ 
  (with cost $\overline{c}(v)$) on the input $P_{V^n}$ 
 of  $(W^{n+}_B, W^{n+}_E)$, where $\overline{c}(v) = \sum_{x\in \cX}c(x)P_{X|V}(x|v)$.
 This principle has enabled us to drastically simplify the performance analysis  of 
 concatenated Poisson wiretap channels.
 

%
\appendices
\section{Proof of Theorem \ref{teiri:han2}\label{appA}}
%

%
 First, set $P_{V^n} = Q$ and  generate a random code 
 $\cC=\{V_1^n, V_2^n, $ $\cdots, V_{M_nL_n}^n\}$ of size $M_nL_n$,  where 
$V_1^n, V_2^n, \cdots, V_{M_nL_n}^n$ are i.i.d. random variables 
with common distribution $Q$ on $\cV^n$, and 
divide the  $M_nL_n$ random codewords 
 $V_1^n, V_2^n, \cdots, $ $V_{M_nL_n}^n$ into $M_n$ subcodes of equal size $L_n$ so that 
 \beqn\label{eq:sub-random}
 \cC_1 & = & \{V_1^n, V_2^n,\cdots, V^n_{L_n}\}, \nonumber\\
 \cC_2 & = & \{V_{L_n+1}^n,  \cdots, V^n_{2L_n}\}, \nonumber\\
\cdots & & \cdots\cdots\cdots\cdots\nonumber\\
\cC_{M_n} & = & \{V_{L_n(M_n-1)+1}^n,  \cdots, V^n_{M_nL_n}\}.
 \eeqn
 In view of (\ref{eq:wag-re1}),  we see that any realization of $X_1^n, X_2^n, \cdots, $ $X_{M_nL_n}^n$ over $\cX^n$
 induced via the auxiliary channel $P_{X^n|V^n}$ by $V_1^n, V_2^n, \cdots, $ $V_{M_nL_n}^n$, respectively, 
 satisfies cost constraint $\Gamma$.
 \\
 \ \
 For each message $i\in \cM_n \equiv \{1,2, \cdots, M_n\}$, the stochastic 
encoder $\varphi_n: \cM_n\to \cV^n$ produces 
the uniform distribution over $\cC_i$.
The decoder $\psi_n^B: \cY^n\to \cM_n$ tries to decode all of these $M_nL_n$ codewords
$V_1^n, V_2^n, \cdots, $ $V_{M_nL_n}^n$.
Then, the {\em reliability} formula for channel $W_B^{n+}:\cV^n\to \cY^n$:
\beq\label{eq:reliability-gallager1}
\Exp_{\cC}[\epsilon_n^B ] \le  \inf_{0\le \rho \le 1}(M_nL_n)^{\rho}e^{-\phi (\rho |W_B^{n+}, Q)}
\eeq
 immediately follows from Gallager \cite{gall} with maximum likelihood decoding, 
 where  $\Exp_{\cC}$ denotes the expectation with respect to 
 the random code $\cC$. 
 \\
 \ \ 
 Next, for each $i\in \cM_n$ we use the subcode $\cC_i$ 
 to produce an output distribution on $\cZ^n$ 
that approximates enough the target output distribution  $\pi_n$ on $\cZ^n$ generated  
via channel $W_E^{n+}:\cV^n\to\cZ^n$
due to the input distribution $P_{V^n}$ (i.e., 
 $\pi_n(\ssz) = \sum_{\ssv\in \cV^n}W_E^{n+}(\ssz|\ssv)P_{V^n}(\ssv)$)
 (the {\em resolvability}
\footnote{Csisz\'ar \cite{csis-all} is the first  who has looked at the secrecy problem with wiretap channels
from the viewpoint of resolvability devised by  Han and Verd\'u \cite{ver-han}.}
 problem).
Let $U_i^n$ be the random variable taking values uniformly in the subcode $\cC_i$, and
let $Z_i^n$ be the output via channel $W_E^{n+}$ due to the input
 $U_i^n$ ($i =1,2,\cdots, M_n$), with the probability distribution of  $Z_i^n$ 
 denoted by $P_n^{(i)}$.
We now evaluate the degree of approximation in terms of the divergence $D(P_n^{(i)}||\pi_n).$
By symmetry of the subcodes, we can focus on the case $i=1$ without loss of generality.
For notational simplicity, with $P_{Z^n} = \pi_n$, set
\[
W^n = W_E^{n+}, \quad i_{V^nW^{n}}(\ssv, \ssz) = \log\frac{W_E^{n+}(\ssz|\ssv)}{P_{Z^n}(\ssz)}.
\]
Then, we have 
\beqn\label{eq:simple-proof1}
\lefteqn{\Exp_{\cC_1}D(P_n^{(1)}||\pi_n)}\nonumber\\
& = & \sum_{\ssz\in \cZ^n}\sum_{\ssc_1\in\cV^n} \cdots  \sum_{\ssc_{L_n} \in \cV^n}
P_{V^n}(\ssc_1)\cdots  P_{V^n}(\ssc_{L_n})\nonumber\\      
& &   \cdot \frac{1}{L_n}\sum_{j=1}^{L_n}W^n(\ssz|\ssc_j)\log 
\left(\frac{1}{L_n} \sum_{k=1}^{L_n}\exp i_{V^nW^n}(\ssc_k, \ssz)\right)\nonumber\\     
   & = & \sum_{\ssc_1\in\cV^n} \cdots  \sum_{\ssc_{L_n} \in \cV^n}
P_{V^n}(\ssc_1)\cdots  P_{V^n}(\ssc_{L_n})\nonumber\\      
& &   \cdot \sum_{\ssz\in \cZ^n}W^n(\ssz|\ssc_1)\log 
\left(\frac{1}{L_n} \sum_{k=1}^{L_n}\exp i_{V^nW^n}(\ssc_k, \ssz)\right)\nonumber\\   
&\le &    \sum_{\ssc_1\in\cV^n}\sum_{\ssz\in \cZ^n} W^n(\ssz|\ssc_1)P_{V^n}(\ssc_1)\nonumber\\     
& & \cdot \log \left(\frac{1}{L_n}\exp i_{V^nW^n} (\ssc_1, \ssz)\right.\nonumber\\
& &  \quad \quad +\left. \frac{1}{L_n}\sum_{k=2}^{L_n}\Exp\exp i_{V^nW^n}(V^n_k, \ssz)\right)\nonumber\\ 
&\le & \Exp \left[\log \left(1+ \frac{1}{L_n}\exp i_{V^nW^n}(V^n, Z^n)\right)\right],
\eeqn
where the first inequality follows from the concavity of the logarithm and the second one is a result of
\[
\Exp [\exp i_{V^nW^n}(V_k^n, \ssz)] =1
\]
for all $\ssz \in \cZ^n$ and $k=1,2,\cdots, L_n$. Now, apply a simple inequality 
with $0<\rho \le 1$  and $x\ge 0$:
\[
\log (1+x) = \frac{\log (1+x)^{\rho}}{\rho}\le \frac{\log (1+x^{\rho})}{\rho}\le \frac{x^{\rho}}{\rho}
\]
to (\ref{eq:simple-proof1}) to eventaully  obtain
\[
\Exp_{\cC_1}D(P_n^{(1)}||\pi_n)
 \le \inf_{0< \rho \le1}\frac{e^{-\psi (\rho |W_E^{n+}, Q)}}{\rho L_n^{\rho}},
 \]
from which it follows that 
\[
\Exp_{\cC}\left[\frac{1}{M_n}\sum_{i=1}^{M_n}D(P_n^{(i)}||\pi_n)\right] 
 \le \inf_{0< \rho \le1}\frac{e^{-\psi (\rho |W_E^{n+}, Q)}}{\rho L_n^{\rho}},
\]
that is,
\beq\label{eq:eventually1}
\Exp_{C}[\delta_n^E]  \le \inf_{0< \rho \le1}\frac{e^{-\psi (\rho |W_E^{n+}, Q)}}{\rho L_n^{\rho}}.
\eeq

Thus, in view of   (\ref{eq:reliability-gallager1}) and (\ref{eq:eventually1}) with Markov inequality,
\footnote{Markov inequality tells that $\Pr\{\epsilon_n^B \le 2\Exp_{C}[\epsilon_n^B]\} >1/2$  and
$\Pr\{\delta_n^E \le 2\Exp_{C}[\delta_n^E]\} >1/2$ Hence, 
$\Pr\{\epsilon_n^B \le 2\Exp_{C}[\epsilon_n^B]$ \mbox{\ and}
$\delta_n^E \le 2\Exp_{C}[\delta_n^E]\}>0.$ This implies that 
there exists at least one realization of $\epsilon_n^B$ and $\delta_n^E$
satisfying (\ref{eq:hana1}) and (\ref{eq:hana2c}).
}
we conclude that there exists at least one non-random pair $(\varphi_n, \psi_n^B)$ of encoder 
(satisfying the cost constraint $\Gamma$) and decoder
satisfying (\ref{eq:hana1}), (\ref{eq:hana2c}).
Moreover, upper bound (\ref{eq:hana2c2}) comes from (\ref{eq:hana2c}) and a simple inequality (due to 
H\"older's inequality):
\beq\label{eq:simple}
\left(\sum_{\ssv}Q(\ssv)W_E^{n+}(\ssz|\ssv)^{1+\rho}\right)W^{n+}_Q(\ssz)^{-\rho}\le 
\left(\sum_{\ssv}Q(\ssv)W_E^{n+}(\ssz|\ssv)^{\frac{1}{1-\rho}}\right)^{1-\rho} 
\eeq
for  $0\le \rho < 1$, 
thereby completing the proof of the theorem. \QED 

\section{Proof of Theorem \ref{teiri:hanw2}\label{appB}}
%
%
%

%
Suppose that $P_X$ satisfies the condition
$\sum_{x \in \cX}P_{X}(x)c(x) \le \Gamma,$
 %
and define
  \beq\label{eq:ransuu1}
  \chi(\ssx) = \left\{
  \begin{array}{cl}
  1 & \mbox{for} \  \sum_{i=1}^n c(x_i) \le n\Gamma,\\
  0 & \mbox{otherwise};
  \end{array}
  \right.
  \eeq
  \beq\label{eq:ransuuji3}
  \mu_n = \sum_{\ssx}\chi(\ssx)\prod_{i=1}^nP_{X}(x_i).
  \eeq
 It is easy to see that $\lim_{n\to\infty}\mu_n= 1$ if 
 $\sum_{x \in \cX}P_{X}(x)c(x) <\Gamma$; and $\lim_{n\to\infty}\mu_n= 1/2$ otherwise (i.e., 
 $\sum_{x \in \cX}P_{X}(x)c(x) =\Gamma$),
 by means of the central limit theorem.
 We rewrite $\mu_n$ as follows:
 \beqn\label{eq;yari1}
 \mu_n &=& P_{X^n}(\cX^n(\Gamma))\nonumber\\
 &=& \sum_{\ssv\in \cV^n}P_{X^n|V^n}(\cX^n(\Gamma)|\ssv)P_{V^n}(\ssv),
 \eeqn
 then, by means of Markov inequality
 \footnote{Markov inequality tells that if $\sum_{v\in \cV}P_{X|V}(\cX|v)P_V(v) \ge 1- \kappa$ then there exist a subset 
 $\cV_0\subset \cV$ such that  $P_V(\cV_0) \ge 1 -\sqrt{\kappa}$ and $P_{X|V}(\cX|v) \ge 1 -\sqrt{\kappa}$
 for all $v \in \cV_0.$}
  there exists a subset $\cT_0 \subset \cV^n$ such that
 \beqn
 \alpha_n \stackrel{\Delta}{=}P_{V^n}(\cT_0) &\ge & 1-\sqrt{1-\mu_n}
 \stackrel{\Delta}{=} \beta_n,\label{eq:ransuujir1}\\
\gamma_n(\ssv) \stackrel{\Delta}{=} P_{X^n|V^n}(\cX^n(\Gamma)|\ssv) &\ge & \beta_n 
 \quad \mbox{for all\ } \ssv\in \cT_0.\label{eq:ransuujir2}
 \eeqn
 Obviously, $\lim_{n\to\infty}\alpha_n= \lim_{n\to\infty}\beta_n=\lim_{n\to\infty}\gamma_n(\ssv)= 1$
 if  $\sum_{x \in \cX}P_{X}(x)c(x) <\Gamma$; otherwise 
 $\lin\alpha_n\ge \lim_{n\to\infty}\beta_n=1-1/\sqrt{2}$ and $\lin\gamma_n(\ssv)\ge 1-1/\sqrt{2}$.
Thus, we can define 
 \beqn\label{eq;yari1}
 \tilde{P}_{V^n}(\ssv) &=& \frac{P_{V^n}(\ssv)}{\alpha_n} \quad  ( \ssv \in \cT_0),\label{eq:ransuujir5}\\
\tilde{P}_{X^n|V^n}(\ssx|\ssv) &=& \frac{P_{X^n|V^n}(\ssx|\ssv)}{\gamma_n(\ssv)}
\quad (\ssx\in \cX^n(\Gamma), \ssv \in \cT_0),\label{eq:ransuujir6}
  \eeqn
  which  obviously  specify a probability distribution and a conditional probability distribution.
On the other hand, notice that $\chi(\ssx)$ can be  upper bounded (for all $\ssx\in \cX^n$) as 
 %
 %
   \beq\label{eq:smooth1}
  \chi(\ssx) \le \exp\left[(1+\rho)r\left(n\Gamma-\sum_{i=1}^nc(x_i)\right)\right],
  \eeq
  where $r\ge 0$ is an arbitrary   number. Now consider $\tilde{P}_{V^n}(\ssv)$ and
  $\tilde{P}_{X^n|V^n}(\ssx|\ssv)$ as $Q(\ssv)$   in (\ref{eq:hana1}) and 
 $P_{X^n|V^n}(\ssx|\ssv)$  in (\ref{eq:chui10}), respectively,  
to obtain
\beq\label{eq:yatto1}
\epsilon_n^B \le  \frac{1}{\alpha _n^{1+\rho}}(M_nL_n)^{\rho}
\sum_{\ssy}\left(\sum_{\ssv}P_{V^n}(\ssv)
W^{n+}_{B}(\ssy|\ssv)^{\frac{1}{1+\rho}}\right)^{1+\rho},\label{eq:hanacgE}
\eeq
and
\beqn
\lefteqn{W^{n+}_B(\ssy|\ssv)}\nonumber\\
 &= &\frac{1}{\gamma_n(\ssv)}\sum_{\ssx \in {\cX^n(\Gamma)}}W^{n}_B(\ssy|\ssx)
 P_{X^n|V^n}(\ssx|\ssv)
\nonumber\\
&=& \frac{1}{\gamma_n(\ssv)}\sum_{\ssx\in \cX^n}W^{n}_B(\ssy|\ssx)\chi(\ssx)
P_{X^n|V^n}(\ssx|\ssv)\nonumber\\
&\le & \frac{1}{\gamma_n(\ssv)}\sum_{\ssx\in \cX^n}W^{n}_B(\ssy|\ssx)\exp\left[(1+\rho)
r\left(n\Gamma -\sum_{i=1}^n c(x_i) \right)
\right] P_{X^n|V^n}(\ssx|\ssv)\nonumber\\
&\le & \frac{1}{\beta_n}\sum_{\ssx\in \cX^n}W^{n}_B(\ssy|\ssx)\exp\left[(1+\rho)
r\left(n\Gamma -\sum_{i=1}^n c(x_i) \right)
\right] P_{X^n|V^n}(\ssx|\ssv),\nonumber\\
& &\label{eq:yatto2}
\eeqn
which together with (\ref{eq:iid-dist1endr1}) $\sim$ (\ref{eq:iid-dist1endr3}) 
yields, with $0\le \rho \le 1$,  
\beqn 
\epsilon_n^B
   & \le &
\frac{2}{\alpha_n^{1+\rho}\beta_n}(M_nL_n)^{\rho}\nonumber\\
 & & \cdot \left[\sum_{y\in \cY}\left(
  \sum_{v\in \cV}q(v)\left[\sum_{x \in \cX}W_B(y|x)
P_{X|V}(x|v)e^{(1+\rho)r[\Gamma -c(x)]}\right]^{\frac{1}{1+\rho}}\right)^{1+\rho}\right]^n, 
\nonumber\\
& & \label{eq:istan1A}
  \eeqn
   Next, let us evaluate upper bound (\ref{eq:hana2c2}).
In the way similar to the argument above with $-\rho$ in place of $\rho$,
we obtain  with $0< \rho <1$:
\beqn\label{eq:delta-wing}
\delta_n^E
 & \le & 
\frac{2}{\alpha_n^{1-\rho}\beta_n}\frac{1}{\rho L_n^{\rho}}\nonumber\\
 & & \cdot \left[\sum_{z\in \cZ}\left(
  \sum_{v\in \cV}q(v)\left[\sum_{x \in \cX}W_E(z|x)
P_{X|V}(x|v)e^{(1-\rho)r[\Gamma -c(x)]}\right]^{\frac{1}{1-\rho}}\right)^{1-\rho}\right]^n.
 \nonumber\\
& & \label{eq:istan2B} \QED
\eeqn
  \section{Proof of Lemma \ref{hodai:wag1q}}\label{appC}
  

We first show Assertion 1).  It follows from (\ref{eq:func11}) that
\beqn
\lefteqn{F_c(q, R_B, R_E,+\infty)}\nonumber\\
&\equiv & \sup_{r\ge 0}\sup_{0\le \rho \le 1}\left(\phi (\rho |W_B, q, r)
- \rho (R_B+R_E)\right).
 \label{eq:func11wag}
\eeqn
Let us first consider the  condition for the $\sup_{0\le \rho \le 1}$ 
on the right-hand side of (\ref{eq:func11wag}) to be attained at $\rho=0$, which is obviously
\beq\label{eq:wakarea1}
\frac{\partial}{\partial \rho}\phi (\rho |W_B, q, r)\Biggl|_{\rho=0} =R_B+R_E.
\eeq
Furthermore, in order to fix the value of $r$ to attain $\sup_{r\ge 0}$ under the condition $\rho=0$, set
\beqn\label{eq:hanten1w}
f(r) & \stackrel{\Delta}{=}&\phi (\rho =0 | W_B, q, r)\nonumber\\
&=& -\log\left[\sum_{y\in \cY}\sum_{v\in \cV}q(v)\sum_{x\in \cX}W_B(y|x)P_{X|V}(x|v)e^{r[\Gamma -c(x)]}\right]
\nonumber\\
& = & -\log\left[\sum_{x\in \cX}P_X(x)e^{r[\Gamma -c(x)]}\right].
\eeqn
Then,  setting
\beqns
A&=& \sum_{x\in \cX}P_X(x)e^{r[\Gamma -c(x)]},\\
B&=&\sum_{x\in \cX}P_X(x)(\Gamma -c(x))e^{r[\Gamma -c(x)]},\\
C&=& \sum_{x\in\cX}P(x)(\Gamma - c(x))^2e^{r[\Gamma -c(x)]},
\eeqns
we have
\beqn\label{eq:wagen1}
f^{\prime}(r)&=& -\frac{B}{A},
%
\eeqn
  \beqn\label{eq:wagen2}
f^{\prime\prime}(r) &=& -\frac{AC -B^2}{A^2} \le 0,
\eeqn
where the last step  follows from Cauchy-Schwarz inequality and this means that $f(r)$ is concave. . 
On the other hand, from (\ref{eq:hanten1w})   
and (\ref{eq:wagen1}) we have
\beqn\label{eq:wagen2wag}
f(0)&=&0,\\
f^{\prime}(0) &=& -\sum_{x\in \cX}P_X(x)(\Gamma-c(x)) \nonumber\\
&=& -\left(\Gamma -\sum_{x\in\cX}P(x)c(x)\right)\le 0,\label{eq:wasurete1}
\eeqn
 where the last step comes from the assumed condition. Therefore, we conclude 
 from (\ref{eq:wagen2}) $\sim$ (\ref{eq:wasurete1}) that $f(r)$ attains the maximum value zero at $r=0$.
 Then, equation (\ref{eq:wakarea1}) reduces to
 \beq
\frac{\partial}{\partial \rho}\phi (\rho |W_B, q, r=0)\Biggl|_{\rho=0} =R_B+R_E.
\eeq     
 On the other hand, it is easy to see that the left-hand side is equal to $I(q, W^+_B)$,
and thus Assertion 1) was proved. In the same way Assertion 2) can also be shown, using, instead of  
(\ref{eq:func11wag}),  
   \beqn
\lefteqn{H_c(q, R_E,+\infty)}\nonumber\\
&\equiv &\sup_{r\ge 0} \sup_{0<\rho < 1}\left(\phi (-\rho |W_E, q, r)
+\rho R_E\right).
 \label{eq:func11wag2}
\eeqn
 Next consider about Assertion 3). In view of the form of the right-hand side of (\ref{eq:func11wag}), 
  we can invoke the same argument as in  Gallager \cite{gall} to conclude that $F_c(q, R_B, R_E,+\infty)$  is 
  monotone strictly decreasing  convex  function of $R_B+R_E$ 
for $R_B+R_E< I(q, W^+_B),$ from which combined with Assertion 1) the positivity follows.
  Similarly for Assertion 4).
  \QED 
  %
  %

%
\section*{Acknowledgments}

The authors are  grateful to Vincent Tan, Ryutaro Matsumoto 
for valuable discussions to improve the earlier manuscript. 
They are also  indebted to Associate Editor and Reviewers for their helpful critical comments 
which have occasioned to make an indeed  major revision of the ealier manuscript.


\begin{thebibliography}{999}
%
%

%
 
 
 


\bibitem{wyner-wire} A. D. Wyner, ``The wire-tap channel," 
 {\em Bell Syst. Tech. J.}, vol.54, pp.1355-1387, 1975


\bibitem{csis-kor-3rd} I. Csisz\'ar and J. K\"orner, ``Broadcast channels with confidential messages,"
 {\em IEEE Transactions  Information Theory}, vol.24, no.3, pp.339-348, 1978


\bibitem{wagner} A. Laourine and A. B. Wagner, ``The degraded Poisson wiretap channel,"
 {\em  IEEE Transactions on Information Theory}, vol.IT-58, no.12, pp.7073-7085, 2012


\bibitem{hayashi-exp} M. Hayashi, ``Exponential decreasing rate of leaked information 
in universal random privacy amplification,"
 {\em  IEEE Transactions on Information Theory}, vol.IT-57, no.6, pp. 3989-4001, 2011
 
 
 

\bibitem{cov-tom} T.M. Cover and J.A. Thomas, {\it Elements of Information Theory,}
2nd ed.,  Wiley, New York, 2006%

\bibitem{hayashi-wire} M. Hayashi, ``General nonasymptotic and asymptotic formulas 
in channel resolvability and identification capacity and their application to the wiretap channel,"
 {\em  IEEE Transactions on Information Theory}, vol.IT-52, no.4, pp. 1562-1575, 2006
 
 

 \bibitem{csis-all} I. Csisz\'ar, ``Almost independence and secrecy capacity," 
 {\em  Problems of Information Transmission}, vol.32, no.1, pp. 40-47, 1996

\bibitem{hou-kra} J. Hou and G. Kramer, ``Effective secrecy: reliability, confusion and stealth," 
{\em  ArX: 1311.1411v3 [cs.IT]}, Jan. 2014

 
 

\bibitem{csis-kor-2nd} I. Csisz\'ar and J. K\"orner, {\em Information Theory: 
Coding Theorems for Discrete Memoryless Systems}, 2nd ed., Cambridge University Press, 2011

\bibitem{pinsker} M.S. Pinsker, {\em Information and Information Stability of Random 
Variables and Processes},
 Holden-Day, San Francisco, 1964

%
\bibitem{gall} R. G. Gallager, {\it Information Theory and Reliable Communication,}
Hoboken, NJ, Wiley, 1968


%
%
\bibitem{han-spec} T. S. Han, {\em Information Spectrum Methods in Information Theory},
Springer, New York, 2003%

%
\bibitem{ver-han} T. S. Han and S. Verd\'u, ``Approximation theory of output statistics,"
 {\em  IEEE Transactions on Information Theory}, vol.IT-399, no.3, pp. 752-772, 1993
 %

\bibitem{bloch} M.R. Bloch and J.N.Laneman, ``Strong secrecy from channel resolvability," 
 {\em  IEEE Transactions on Information Theory}, vol.IT-59, no.12, pp. 8077-8098, 2013
 
 

 %

 
%
\bibitem{wyner1} A. D. Wyner, ``Capacity and error exponent
for the direct detection photon channel--Part I,"
 {\em  IEEE Transactions on Information Theory}, vol.IT-34, no.6, pp.1449-1461, 1988
 
%
 
%
%
\bibitem{lapidoth} A. Lapidoth, E. Telater and R. Urbanke, ``On wide-band broadcast channels,"
 {\em  IEEE Transactions on Information Theory}, vol.IT-49, no.12, pp. 3250-3258, 2003
 
%
 
\bibitem{hellman} S.K.Lueng-Yan-Cheong and M.Hellamn, ``The Gaussian wire-tap channel,"
 {\em  IEEE Transactions on Information Theory}, vol.IT-24, no.4, pp.451-456, 1978
 
\bibitem{vis} Tie Liu and P. Vithwanath, ``An extremal inequality motivated by multi terminal
information-theoretic problems,"
 {\em  IEEE Transactions on Information Theory}, vol.IT-53, no.5, pp.1839-1851, 2007


\end{thebibliography}
\end{document}